\documentclass[aps,preprintnumbers,eqsecnum,amsmath,amssymb,nofootinbib]{revtex4}  %eqsecnum,
\usepackage{graphicx} 
\usepackage{bm}
\usepackage{slashed}
\renewcommand\theequation{\arabic{section}.\arabic{equation}}
\begin{document}
\title{Charming-loop contribution to $B_s\to \gamma\gamma$ decay}
\author{Ilia Belov$^{a}$, Alexander Berezhnoy$^b$, and Dmitri Melikhov$^{b,c,d}$}
\affiliation{
$^a$INFN, Sezione di Genova, via Dodecaneso 33, I-16146 Genova, Italy\\
$^b$D.~V.~Skobeltsyn Institute of Nuclear Physics, M.~V.~Lomonosov Moscow State University, 119991, Moscow, Russia\\
$^c$Joint Institute for Nuclear Research, 141980 Dubna, Russia\\
$^d$Faculty of Physics, University of Vienna, Boltzmanngasse 5, A-1090 Vienna, Austria}

\begin{abstract}
  We present a detailed theoretical study of nonfactorizable contributions of the charm-quark loop to the
  amplitude of the $B_s\to \gamma\,\gamma$ decay. This contribution involves
  the $B$-meson three-particle Bethe-Salpeter amplitude, $\langle 0|\bar s(y)G_{\mu\nu}(x)b(0)|\bar B_s(p)\rangle$,  
  for which we take into account constraints from analyticity and continuity. 
 The charming-loop contribution of interest may be described as a correction to the Wilson coefficient $C_{7\gamma}$, 
 $C_{7\gamma}\to C_{7\gamma}(1+\delta C_{7\gamma})$. We calculate an explicit dependence of $\delta C_{7\gamma}$ on the
 parameter $\lambda_{B_s}$. Taking into account all theoretical uncertainties, $\delta C_{7\gamma}$ may be predicted
 with better than 10\% accuracy for any given value of $\lambda_{B_s}$. For our benchmark point $\lambda_{B_s}=0.45$ GeV,
 we obtain $\delta C_{7\gamma}=0.045\pm 0.004$. Presently, $\lambda_{B_s}$ is not known with high accuracy, but its
 value is expected to lie in the range $0.3\le \lambda_{B_s}({\rm GeV})\le 0.6$.
 The corresponding range of $\delta C_{7\gamma}$ is found to be $0.02\le \delta C_{7\gamma}\le 0.1$.
One therefore expects the correction given by charming loops at the level of at least a few percent. 
\end{abstract}
\date{\today}
\maketitle
\normalsize

\section{Introduction}
\label{Sec_introduction}
Charming loops in rare flavour-changing neutral current (FCNC) decays of the $B$-meson 
have impact on the $B$-decay observables \cite{neubert} and provides an unpleasant noise for
the studies of possible new physics effects (see, e.g., 
\cite{ciuchini2022,ciuchini2020,ciuchini2021,diego2021,matias2022,gubernari2022,hurth2022,stangl2022,diego2023a,diego2023b}).

A number of theoretical analyses of nonfactorizable (NF) charming loops in FCNC $B$-decays has
been published: In \cite{voloshin}, an effective gluon-photon local operator describing the charm-quark loop has been 
calculated as an expansion in inverse charm-quark mass $m_c$ and applied to inclusive $B\to X_s\gamma$ decays
(see also \cite{ligeti,buchalla}); in \cite{khod1997}, NF corrections in $B\to K^*\gamma$ using local
operator product expansion (OPE) have been studied;
NF corrections induced by the {\it local} photon-gluon operator 
have been calculated in \cite{zwicky1,zwicky2} in terms of the light-cone (LC) 3-particle antiquark-quark-gluon
Bethe-Salpeter amplitude (3BS) of $K^*$-meson \cite{braun,ball1,ball2} with two field operators having equal coordinates,  
$\langle 0| \bar s(0)G_{\mu\nu}(0) u(x)|K^*(p)\rangle$, $x^2=0$. Local OPE for the charm-quark loop in FCNC $B$ decays
leads to a power series in $\Lambda_{\rm QCD} m_b/m_c^2\simeq 1$.
To sum up $O(\Lambda_{\rm QCD} m_b/m_c^2)^n$ corrections, Ref.~\cite{hidr} obtained a {\it nonlocal} photon-gluon
operator describing the charm-quark loop and evaluated its effect making use of 3BS of the $B$-meson
in a {\it collinear} LC configuration $\langle 0| \bar s(x)G_{\mu\nu}(ux) b(0)|\bar B_s(p)\rangle$, $x^2=0$ \cite{japan,braun2017}.
The same collinear approximation [known to provide the dominant 3BS contribution to meson tree-level
form factors \cite{braun1994,offen2007}] was applied also to the analysis of other FCNC $B$-decays \cite{gubernari2020}.

In later publications \cite{mk2018,m2019,m2022,m2023}, it was demonstrated that the dominant contribution to
FCNC $B$-decay amplitudes is actually given by the convolution of a hard kernel with the 3BS in 
a different configuration --- a {\it double-collinear} light-cone configuration
$\langle 0| \bar s(y)G_{\mu\nu}(x) b(0)|\bar B_s(p)\rangle$, $y^2=0$, $x^2=0$, but $x\,y \ne 0$. The corresponding 
factorization formula was derived in \cite{m2023}.
The first application of a double-collinear 3BS to FCNC $B$-decays was presented in \cite{wang2022}.

In this paper, we study NF charming loops in $B_s\to \gamma\gamma$ decays making use of the generic 3BS of the $B$-meson.
The main new features of this paper compared to the previous analyses, in particular to \cite{wang2022}, are as follows:

\noindent
(i) The generic 3BS of the $B$-meson contains new Lorentz structures (compared to the collinear and the double-collinear
approximations) and new three-particle distribution amplitudes (3DAs) that appear as the coefficients
multiplying these Lorentz structures. Analyticity and continuity of the 3BS as the function of its arguments at the point
$xp=yp=x^2=y^2=0$ leads to certain constraints on the 3DAs \cite{m2023} which we take into account. 

\noindent
(ii) We derive the convolution formulas for the $B_s\to\gamma^*\gamma^*$ form factors involving this generic 3BS,
and obtain the corresponding numerical predictions. We check that the deviation between our analysis
and the analysis based on double-collinear 3BS differ by $O(\lambda_{B_s}/M_B)$ terms that in practical calculations
give a $\sim$20\% difference. 

\noindent
The paper is organized as follows: Section \ref{sec:Heff} presents general formulas for the top and charm contribution to the
$B_s\to\gamma\gamma$ amplitude. Section  \ref{sec:loop} considers the $\langle AVV\rangle$ charm-quark triangle and gives a
convenient representation for this quantity via the gluon field strength $G_{\mu\nu}$ merely 
(not involving $A_\mu$ itself). In Section  \ref{sec:3BS}, properties of the 3BS of the $B$-meson in the general
noncollinear kinematics are discussed and properly modified 3DAs are constructed. Section  \ref{sec:5} presents
the numerical results for the form factors and for the nonfactorizable charm-loop correction to the
$B_s\to\gamma\gamma$ amplitude. Section 6 gives our concluding remarks. Appendix \ref{sec:appA} compares the definition 
of the amplitude adopted in this paper with the one of \cite{mnk2018}. Appendix \ref{Appendix_Anathomy} gives details of the
numerical results for the form factors. 

%\newpage
%%%%%%%%%%%%%%%%%%%%%%%%%%%%%%%%%%%%%%%%%%%%%%%%%%%%%%%%%%%%
%  SECTION II
%%%%%%%%%%%%%%%%%%%%%%%%%%%%%%%%%%%%%%%%%%%%%%%%%%%%%%%%%%%%%
\section{Top and charm contributions to $B_s\to \gamma\gamma$}
\label{sec:Heff}
\subsection{The $b\to d,s$ effective Hamiltonian}
A standard theoretical framework for treating FCNC $b\to q$ ($q=s,d$) transitions is provided by the Wilson OPE: 
the $b\to q$ effective Hamiltonian describing dynamics at the scale $\mu$, appropriate for $B$-decays, reads 
\cite{Grinstein:1988me,Burasa,Burasb}: 
\begin{eqnarray}
\label{Heff}
H_{\rm eff}^{b\to q}=\frac{G_F}{\sqrt{2}}V^*_{tq}V_{tb}\sum_i C_i(\mu) {\cal O}_i^{b\to q}(\mu),  
\end{eqnarray}
$G_F$ is the Fermi constant and $V_{ij}$ are CKM matrix elements.
The SM Wilson coefficients relevant for our analysis at the scale $\mu_0=5$ GeV have
the following values [corresponding to $C_2(M_W)=-1$]: 
$C_1(\mu_0)=0.241$, $C_2(\mu_0)=-1.1$, $C_7(\mu_0)=0.312$ \cite{Burasa,Burasb,Simulaa,Simulab,hidr}. 

The basis operators ${\cal O}_i^{b\to q}(\mu)$ contain only light degrees of freedom   
($u$, $d$, $s$, $c$, and $b$-quarks, leptons, photons and gluons); the heavy degrees of freedom of the 
SM ($W$, $Z$, and $t$-quark) are integrated out and their contributions are encoded in the Wilson coefficients $C_i(\mu)$. 
The light degrees of freedom remain dynamical and 
the corresponding diagrams containing these particles in the loops -- in the case of our interest virtual $c$ quarks -- 
should be calculated and added to the diagrams generated by the effective Hamiltonian. 

%%%%%%%%%%%%%%%%%%%%%%%%%%%%%%%%%%%%%%%%%%%%%%%%%%%%
\subsection{The penguin contribution}
The top-quark contribution to $B_s\to\gamma\gamma$ decay is generated by penguin operator in (\ref{Heff})\footnote{
Our notations and conventions are: 
$\gamma^5=i\gamma^0\gamma^1\gamma^2\gamma^3$, 
$\sigma_{\mu\nu}=\frac{i}{2}[\gamma_{\mu},\gamma_{\nu}]$, 
$\varepsilon^{0123}=-1$, $\epsilon_{abcd}\equiv
\epsilon_{\alpha\beta\mu\nu}a^\alpha b^\beta c^\mu d^\nu$, 
$e=\sqrt{4\pi\alpha_{\rm em}}$. }
\begin{equation}
\label{b2qgamma} 
H_{\text{eff}}^{b\to s\,\gamma} =  \frac{G_F}{\sqrt{2}}  V_{tb}V_{ts}^*\,C_{7\gamma}(\mu){\cal O}_{7\gamma},\qquad 
 {\cal O}_{7\gamma}=-\frac{e}{8\pi^2}m_b \cdot \bar{s}\sigma_{\mu\nu}(1+\gamma_5)b\cdot F^{\mu\nu}.
\end{equation}
The sign of the $b\to d\gamma$ effective Hamiltonian (\ref{b2qgamma})
correlates with the sign of the electromagnetic vertex. 
For a fermion with the electric charge $Q_qe$, we use in the Feynman diagrams the vertex  
\begin{equation}
  iQ_q e \bar q\gamma_\mu q \epsilon^\mu, 
\end{equation}
corresponding to the definition of the covariant derivative in the form
$D_{\mu} = \partial_{\mu} - ie\,Q_q A_{\mu}$. 

The amplitude of the $B\to\gamma\gamma$ transition is defined according to \cite{mn2004,mnk2018}:
\begin{eqnarray}
\label{Atop}
      {\cal A}^{(B\to\gamma\gamma)}_{\text{top}}
      &\equiv&\langle \gamma(q,\varepsilon),\gamma(q',\varepsilon')|H^{b\to s\gamma}_{\rm eff}|\bar B_s(p)\rangle\nonumber\\
&=&-2\,\frac{G_F}{\sqrt{2}}V_{tb}V_{ts}^*\,\frac{e^2}{8\pi^2}\,2m_b C_{7\gamma}(\mu)
\Big[F_{TV}\, \epsilon_{\alpha \alpha' q q'} - i F_{TA}\,(g_{\alpha\alpha'}q'q-q'_{\alpha} q_{\alpha'}) \Big]
\varepsilon_\alpha\varepsilon'_{\alpha'}.
\end{eqnarray}
Here $q,q'$ and $\varepsilon,\varepsilon'$ are momenta and polarization vectors of the outgoing real photons, and 
$F_{TA}$ and $F_{TV}$ are the form factors $F_{TA}(q^2=0,q'^2=0)$ and $F_{TV}(q^2=0,q'^2=0)$.
The latter are defined as \cite{mk2003,mn2004,mnk2018}:
\begin{eqnarray}
\label{ffsmk}
&&\langle\gamma(q',\varepsilon')|\bar s \sigma_{\mu\nu}\gamma_5 b|\bar B_s(p)\rangle\, q^\nu= 
e\,\varepsilon'_{\alpha}\,\left(g_{\mu\alpha}\,q'q- q_{\alpha}q'_{\mu}\right)\, F_{TA}(q^2, q'^2), 
\\
&&\langle\gamma(q',\varepsilon')|\bar s \sigma_{\mu\nu} b|\bar B_s(p)\rangle\, q^{\nu}
=
i\, e\,\varepsilon'_{\alpha}\epsilon_{\mu\alpha q q'}F_{TV}(q^2, q'^2),
\end{eqnarray}
and satisfy a rigorous constraint $F_{TA}(q^2,0)=F_{TV}(q^2,0)$. Notice that the strange-quark charge $Q_s$ (or $Q_b$
in the $1/m_b$-subleading diagram where the photon is emitted by the $b$-quark) is included in the form factors
$F_{TA}$ and $F_{TV}$ \cite{mnk2018}.

%\newpage
\subsection{Nonfactorizable charm-quark loop correction to $B_s\to \gamma\gamma$}

As already noticed, the light degrees of freedom remain dynamical and their contributions
should be taken into account separately. The relevant terms in $H_{\rm eff}^{b\to s}$ are those containing
four-quark operators: 
\begin{equation}
\label{b2qO12}
H_{\rm eff}^{b\to s\bar cc} = - \frac{G_{F}}{\sqrt2}\, V_{cb}V^*_{cs}\, 
\left\{C_{1}(\mu){\cal O}_1 + C_{2}(\mu){\cal O}_2\right\}
\end{equation}
where 
\begin{align}
&{\cal O}_1 = \left(\bar s^i\gamma_{\mu}(1-\gamma^5)c^j\right)\; \left(\bar c^k\gamma^{\mu}(1-\gamma^5)b^l \right)\; \delta_{il} \delta_{kj},  \\
&{\cal O}_2 = \left(\bar s^i\gamma_{\mu}(1-\gamma^5)c^j\right)\; \left(\bar c^k\gamma^{\mu}(1-\gamma^5)b^l \right)\; \delta_{ij} \delta_{kl},
\end{align}
differing from each other in the way color indices $i,j,k,l$ are contracted. By Fierz transformation
${\cal O}_2$ may be written in the following form (for anticommuting spinor fields):
%containing color singlet-singlet and octet-octet parts 
\begin{equation}
{\cal O}_2  =
\left(\bar s^i\gamma_{\mu} (1-\gamma^5) b^l\right)\left(\bar c^k\gamma^{\mu}(1-\gamma^5) c^j\right) \left(2 t^a_{il}t^a_{jk}  + \frac{1}{3} \delta_{il} \delta_{jk}\right),
\end{equation}
The singlet-singlet operator ${\cal O}_1$ and the singlet-singlet part of ${\cal O}_2$ at the leading order
generate factorizable charm contributions to the $B\to \gamma\gamma$ amplitude.
These factorizable contributions vanish for real photons in the final state.
A nonzero contribution is induced by the octet-octet part of the operator ${\cal O}_2$ and
needs the emission of one soft gluon from the charm-quark loop. So relevant for us is the octet-octet operator 
\begin{equation}
  \label{Heff8times8}
  H_{\text{eff}}^{b\to s\bar c c[8\times 8]} =
  -\frac{G_F}{\sqrt{2}} V_{cb}V^*_{cs} 2C_2
  \left(\overline{s}\gamma_{\mu}(1-\gamma_5)t^a b\right)\left(\overline{c}\gamma^{\mu}(1-\gamma_5)t^a c\right).
\end{equation}
Therefore, similar to the top contribution, we find 
\begin{eqnarray}
\label{Acharm}
{\cal A}^{(B\to\gamma\gamma)}_{\text{charm}}=
\langle \gamma(q,\varepsilon),\gamma(q',\varepsilon')|H^{b\to s\bar cc[8\times 8]}_{\rm eff}|\bar B_s(p)\rangle. 
\end{eqnarray}
Here quark fields are understood as Heisenberg field operators with respect to the SM interactions. 
Expanding them to the second order in electromagnetic interaction and to the first order in strong interaction gives
\begin{eqnarray}
\label{Acharm1}
&&  {\cal A}^{(B\to\gamma\gamma)}_{\text{charm}}=
  \frac{1}{2}i^3\langle \gamma(q,\varepsilon),\gamma(q',\varepsilon')|\nonumber
  \\
  &&\hspace{1cm}
  \times T\Big\{H^{b\to s\bar cc[8\times 8]}_{\rm eff}(0),\int dz j^{\rm e.m.}_\rho(z)A_\rho(z),
  \int dy j^{\rm e.m.}_\eta(y)A_\eta(y),
  \int dx \bar c(x) \gamma_\nu t^b c(x) g_s B_\nu^b(x)\Big\}|\bar B_s(p)\rangle
\end{eqnarray}
where quark electormagnetic current has the form $j^{e.m.}_\alpha=e\sum Q_i\bar q_i\gamma_\alpha q_i$. 
\begin{figure}[!b]
\begin{center}
\includegraphics[height=5cm]{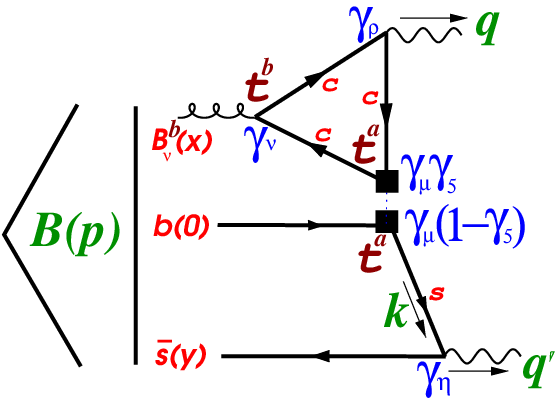}
\caption{\label{Fig:1}One of the diagrams describing charming loop contribution to $B_s\to \gamma\,\gamma$ decay
  via nonfactorizable soft gluon exchange. Other diagrams are those corresponding to an opposite direction
  in the charm-quark loop and diagrams with the interchanged photons 
  $q\leftrightarrow q',\varepsilon\leftrightarrow \varepsilon'$. }
\label{fig:nf}
\end {center}
\end{figure}
Eq.~(\ref{Acharm1}) may be rewritten as:
\begin{eqnarray}
  \label{Acharm2}
{\cal A}^{(B\to\gamma\gamma)}_{\text{charm}}&=&-i^3\frac{G_F}{\sqrt2}2C_2 V_{cb}V_{cs}^*e^2 Q_c Q_s
\int dz dx dy
\left[\varepsilon_\rho e^{iqz}\varepsilon'_{\eta}e^{iq'y}
  +(q\leftrightarrow q',\varepsilon\leftrightarrow \varepsilon')\right]
 \nonumber\\
 &&\times \langle 0|T\Big\{\bar c\gamma_\rho c(z),
  \bar c(0)t^a\gamma_\mu(1-\gamma_5)c(0),
  \bar c(x)t^b\gamma_\nu c(x)\Big\}|0\rangle\nonumber\\
  &&\times \langle 0|T\Big\{\bar s(y)\gamma_\eta s(y),
  \bar s(0)t^a\gamma_\mu (1-\gamma_5)b(0)g_s B^b_\nu(x)\Big\}|\bar B_s(p)\rangle.
\end{eqnarray}
Fig.~\ref{fig:nf} shows one of the corresponding diagrams when the photon is emitted by the $B$-meson valence $s$-quark. 
We will neglect the $1/m_b$-suppressed contribution when the photon is emitted by the valence $b$-quark.

\newpage
A detailed treatment of the operator, describing charm-quark triangle [second line in Eq.~(\ref{Acharm2})]
is given in the next Section.
Here we only notice two important features of this operator: 
\begin{itemize}
  \item 
    The $\bar c\gamma_\mu c$ part of the $V-A$ weak current does not contribute and one is left with
    $\langle VVA\rangle$ charm-quark triangle.
\item
  The $\langle VVA\rangle$ charm-quark triangle contracted with the gluon field $B^b_{\nu}$
  may be written as a gauge-invariant nonlocal operator containing gluon field strength $G^b_{\nu\alpha}$
  for any gluon momentum (cf.~\cite{buchalla}). 
\end{itemize}
Making use of the result for the charm-quark $VVA$ triangle from the next Section,
we obtain the following expression for the amplitude: 
\begin{eqnarray}
\label{Acharm3a}
A^{B\to\gamma\gamma}_{\rm charm}&=&\frac{G_F}{\sqrt{2}}4C_2 V_{cb}V^*_{cs}e^2 Q_sQ_c\,A_{\rho\eta}(q,q')\varepsilon_\rho\varepsilon'_\eta,
\\
A_{\rho\eta}(q,q')&=&\frac{1}{(2\pi)^8}\int dk dy e^{-i(k-q')y} dx d\kappa e^{-i\kappa x} 
\Gamma_{cc}^{\mu\nu\rho(ab)}(\kappa,q)
\langle 0|\bar s(y)\gamma^\eta\frac{\slashed  k+m_s}{m_s^2-k^2}\gamma^\mu(1-\gamma^5) t^a
       B^b_{\nu}(x)
        b(0)|\bar B_s(p)\rangle\nonumber\\
&=&\frac{1}{4(2\pi)^8}\int dk dy e^{-i(k-q')y} dx d\kappa e^{-i\kappa x} 
\overline{\Gamma}_{cc}^{\mu\nu\rho\alpha}(\kappa,q)
\langle 0|\bar s(y)\gamma^\eta\frac{\slashed  k+m_s}{m_s^2-k^2}\gamma^\mu(1-\gamma^5) t^b
        G^b_{\nu\alpha}(x)b(0)|\bar B_s(p)\rangle.
       \nonumber\\
\label{Acharm3b}        
\end{eqnarray}
For real photons in the final state, the amplitude $A_{\rho\eta}(q,q')$ has the same Lorentz structure as
the penguin amplitude and contains two form factors $H_V=H_V(q^2=0,q'^2=0)$ and $H_A=H_A(q^2=0,q'^2=0)$
(Appendix \ref{sec:appA} presents comparison with the form factors defined in \cite{mnk2018}): 
\begin{equation}
  \label{Acharm4}
  A_{\rho\eta}(q,q')= 
  H_V \epsilon_{\rho\eta q q'} - i H_A \left(g_{\rho\eta}\, q q' - q'_{\rho}q_{\eta}\right),
\end{equation}
Comparing Eqs.~(\ref{Atop}) and (\ref{Acharm4}), and taking into account that
$V_{tb}V^*_{ts}\simeq -V_{cb}V^*_{cs}$, 
it is convenient to desctribe the effect of charm as an additions to the
Wilson coefficient $C_{7\gamma}$ (however, non-universal, i.e.
different in the axial and vector Lorentz structures):
\begin{eqnarray}
\epsilon_{\rho\eta qq'}: && C_{7\gamma}\to C_{7\gamma}(1+\delta_V C_{7\gamma}),\nonumber\\
g_{\rho\eta}\, q q' - q'_{\rho}q_{\eta}: && C_{7\gamma}\to C_{7\gamma}(1+\delta_A C_{7\gamma}),
\end{eqnarray}
with 
\begin{eqnarray}
\label{deltaC7}
\delta_{V(A)}C_{7\gamma}=8\pi^2
\,Q_s Q_c \frac{C_2}{C_{7\gamma}}\frac{H_{V(A)}}{m_b\,F_{TV(TA)}}.
\end{eqnarray}
Our goal will be to calculate these corrections.

The $B$-meson structure contributes to the
${\cal A}^{B\to\gamma\gamma}_{\rm charm}$ amplitude via the full set of 3BS 
\begin{eqnarray}
\langle 0|\bar s(y)\Gamma_i t^a b(0) G^a_{\nu\alpha}(x)|\bar B_s(p)\rangle,
\end{eqnarray}
with $\Gamma_i$ the appropriate combinations of $\gamma$-matrices. 
This quantity is not gauge invariant, since it contains field operators at different locations. 
To make it gauge-invariant, one needs to insert Wilson lines between the field operators.
To simplify the full consideration, it is convenient to work in a fixed-point gauge,
where the Wilson lines reduce to unity factors.
As first noticed in \cite{mk2018}, the dominant contribution of charm to amplitudes of FCNC $B$ decays 
comes from the ``double collinear'' LC configuration \cite{m2023},
where $x^2=0$, $y^2=0$, but $xy\ne 0$, i.e. 4-vectors $x$ and $y$ are not collinear. 
Respectively, we need to parametrize the 3BS in this kinematics; this is discussed in Sect.~\ref{sec:3BS}.
But before studying 3BS, we present in the next Section a convenient representation for the
operator describing the contribution of charm-quark loop.

%%%%%%%%%%%%%%%%%%%%%%%%%%%%%%%%%%%%%%%%%%%%%%%%%%%%%%%%%%%%%%%%%%%%%
%   SECTION <VVA>
%%%%%%%%%%%%%%%%%%%%%%%%%%%%%%%%%%%%%%%%%%%%%%%%%%%%%%%%%%%%%%%%%%%%%%
%\newpage
%%%%%%%%%%%%%%%%%%%%%%%%%%%%%%%%%%%%%%%%%%%%%%%%%%%%%%%%%%%%%%%%%%%%%%

\section{Charm-quark $\langle VVA\rangle$ triangle}
\label{sec:loop}
\begin{figure}[!b]
\begin{center}
\includegraphics[height=3cm]{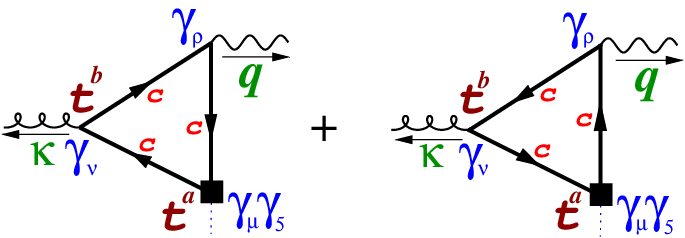}
\caption{The $\langle VVA\rangle$ triangle one-loop diagrams for $\Gamma_{cc}^{\mu\nu\rho\,(ab)}$.}
\label{fig:loop}
\end{center}
\end{figure}

The charm-quark loop contribution is described by the three-point function (see Fig.~\ref{fig:loop}): 
\begin{equation}
\label{gamma_def}
\Gamma_{cc}^{\mu\nu\rho\,(ab)}\left(\kappa,q\right) = \int dx' dz \: e^{i q z+i \kappa x'}
\langle 0 |T\lbrace\bar c(z)\gamma^{\rho}c(z), \bar c(0)\gamma^{\mu} (1-\gamma_5)t^a c(0),
\bar c(x')\gamma^\nu t^b c(x')\rbrace|0 \rangle=\frac12\delta^{ab}\,\Gamma_{cc}^{\mu\nu\rho}(\kappa ,q), 
\end{equation}
where $q$ is the momentum of the external virtual photon (vertex containing index $\rho$) 
and $\kappa$ is the gluon momentum (vertex containing index $\nu$). Here $t^c$, $c=1,\dots,8$ are 
$SU_c(3)$ generators normalized as ${\rm Tr}(t^at^b)=\frac12\delta^{ab}$. 
The octet current $\bar c(0) \gamma^{\mu}(1-\gamma_5)t^a c(0)$ is a charm-quark part of the
octet-octet weak Hamiltonian. Its vector piece does not contribute to $\Gamma_{cc}^{\mu\nu\rho\,(ab)}$
(Furry theorem) and will be omitted.
Taking into account vector-current conservation,
it is convenient to parametrize $\Gamma_{cc}^{\mu\nu\rho}(\kappa ,q)$ as follows \cite{lm} 
\begin{equation}
\label{cquarkloop}
 \Gamma_{cc}^{\mu\nu\rho}(\kappa ,q) =
-i\left(\kappa ^{\mu }+q^{\mu }\right) \epsilon^{\nu\rho\kappa q}\,F_0
-i\left(q^2      \epsilon^{\mu\nu\rho\kappa}-   q^{\rho}\epsilon^{\mu\nu q\kappa}\right)\,F_1 
-i\left(\kappa^2 \epsilon^{\mu\rho\nu q} -\kappa^{\nu} \epsilon^{\mu\rho\kappa q}\right)\, F_2 . 
\end{equation}
The form factors $F_{0,1,2}$ are functions of three independent invariant variables $q^2$, $\kappa^2$, and $\kappa q$.
The lowest order QCD diagrams describing $\Gamma_{cc}^{\mu\nu\rho}(\kappa,q)$
are shown in Fig.~\ref{fig:loop}. A convenient representation of the form factors has the form \cite{max}
\begin{eqnarray}
\label{ffs}
F_i\left(\kappa^2,\kappa q, q^2\right) &=&
\frac{1}{\pi^2}\int\limits_{0}^1 d\xi \int\limits_{0}^{1-\xi}d\eta
\frac{\Delta_i(\xi,\eta)}{m_c^2 - 2\,\xi\eta\:\kappa q - \xi(1-\xi)q^2 - \eta(1-\eta)\kappa ^2},\quad i=0,1,2,
\nonumber
\\
&&\Delta_0 = -\xi\eta, \quad \Delta_1 = \xi(1-\eta-\xi), \quad \Delta_2 = \eta(1-\eta-\xi). 
\end{eqnarray}
This representation may be applied to the physical amplitude in the region of the external momenta far below the thresholds, 
$q^2,\kappa^2,(\kappa+q)^2\ll 4m_c^2$. Taking into account the momentum distribution of quarks and gluons inside
the $B$-meson, the dominant contribution of the charm-quark loop to the $B$-decay amplitude comes from the region
$\kappa^2\sim\Lambda^2_{\rm QCD}$, $(q+\kappa)^2<0$. So, the representation (\ref{ffs})
is applicable and proves convenient for numerical calculations. 

As the next step, one takes the convolution of the amplitude (\ref{cquarkloop}) with the gluon field $B_\nu(x)$.
The Lorentz structures multiplying $F_0$ and $F_1$ contain $\epsilon_{\nu\kappa\alpha_1\alpha_2}$
with some indices $\alpha_1$ and $\alpha_2$. 
After multiplying by $B_\nu(x)$ and performing parts integration,
their contribution may be reduced to the convolution with the gluon strength tensor $G_{\alpha\nu}$: 
\begin{eqnarray}
  \int e^{-i \kappa x}\epsilon^{\alpha\nu\rho\omega}\kappa _{\alpha}B_\nu(x) dx
  &=&
 i \int\left[ \frac{\partial}{\partial x_\alpha}e^{-i \kappa x}\right]\epsilon^{\alpha\nu\rho\omega}B_\nu(x) dx
=
  -\frac{i}{2} \int e^{-i \kappa x}\epsilon^{\alpha\nu\rho\omega}
  \left[\frac{\partial}{\partial x_\alpha}B_\nu(x)-\frac{\partial}{\partial x_\nu}B_\alpha(x)\right] dx\nonumber\\
  &=&
  -\frac{i}{2} \int e^{-i \kappa x}\epsilon^{\alpha\nu\rho\omega} G_{\alpha\nu}(x)dx.
\label{eq:AtoG}
\end{eqnarray}
The Lorentz structure multiplying $F_2$ at first glance does not have this property. However, using the identity 
\begin{eqnarray}
  g^{\alpha_1\alpha_2}\epsilon^{\alpha_3\alpha_4\alpha_5\alpha_6}
 -g^{\alpha_1\alpha_3}\epsilon^{\alpha_2\alpha_4\alpha_5\alpha_6}
 +g^{\alpha_1\alpha_4}\epsilon^{\alpha_2\alpha_3\alpha_5\alpha_6}
 -g^{\alpha_1\alpha_5}\epsilon^{\alpha_2\alpha_3\alpha_4\alpha_6}
+g^{\alpha_1\alpha_6}\epsilon^{\alpha_2\alpha_3\alpha_4\alpha_5}=0,  
\end{eqnarray}
multiplying it by $\kappa_{\alpha_1}\kappa_{\alpha_2}q_{\alpha_6}$, and setting
$\alpha_3\to\mu$, $\alpha_4\to\nu$, $\alpha_5\to\rho$,
this Lorentz structure takes the form
\begin{eqnarray}
\kappa^2 \epsilon^{\mu\nu\rho q}+\kappa^\nu\epsilon^{\mu\rho\kappa q}
=\kappa^\mu \epsilon^{\kappa\nu\rho q}+\kappa^\rho\epsilon^{\kappa\mu\nu q}-\kappa q \epsilon^{\kappa\mu\nu\rho}
\end{eqnarray}
and may be also reduced to the convolution with $G_{\alpha\nu}$.
Finally, the operator describing the
contribution of the charm-quark loop takes the form
\begin{eqnarray}
  \int d\kappa  e^{-i\kappa x~}\Gamma_{cc}^{\mu\nu\rho\,(ab)}(\kappa ,q)B^b_{\nu}(x)dx =
 \frac{1}{4}\int d\kappa  e^{-i\kappa x~}{\overline\Gamma}_{cc}^{\mu\nu\rho\alpha} (\kappa, q) G^a_{\nu\alpha}(x)dx
\end{eqnarray}
with
\begin{eqnarray}
{\overline\Gamma}_{cc}^{\mu\nu\rho\alpha} (\kappa, q)=
\left(\kappa ^{\mu }+q^{\mu }\right)\epsilon^{\nu\rho\alpha q}\,F_0  
+\left(q^{\rho}\epsilon^{\mu\nu\alpha q} + q^2\epsilon^{\mu\nu\rho\alpha}\right)\,F_1
+\left(\kappa^{\mu}\epsilon^{\alpha\nu\rho q} +
\kappa ^{\rho}\epsilon^{\alpha\mu\nu q}-\kappa q\, \epsilon^{\alpha\mu\nu\rho}\right)\,F_2.     
\end{eqnarray}
${\overline\Gamma}_{cc}^{\mu\nu\rho\alpha}(\kappa,q)$ is real in the Euclidean region and 
$\Gamma_{cc}^{\mu\nu\rho} (\kappa, q)=-i \,{\overline\Gamma}_{cc}^{\mu\nu\rho\alpha} (\kappa, q)\kappa_\alpha$.
%%%%%%%%%%%%%%%%%%%%%%%%%%%%%%%%%%%%%%%%%%%%%%%%%%%%%%%%%%%%%%%%%%%%%%%%%%
%\newpage

%********************** SECT 4 ************************

\section{3BS of the $B$-meson in a noncollinear kinematics
\label{sec:3BS}}
As already mentioned, the contribution of collinear LC configuration dominates the 3BS corrections to the
$B\to \pi,K$ form factors. These corrections reflect the following picture: in the rest frame of the $B$-meson, 
a fast light quark, produced in weak decay of an almost resting $b$-quark, 
emits a soft gluon and continues to move practically in the same direction, before it fragments into the final light meson. 

Contributions of charming loops in FCNC $B$-decay have a qualitatively different picture \cite{m2022}:
In the rest frame of the decaying $B$-meson,
two fast systems produced in the weak decay of an almost resting $b$-quark move in opposite space directions.
Formulated in terms of the LC variable, this means that the $s$-quark produced in weak decay
moves along one of the LC directions, whereas the $\bar cc$-pair moves along the other LC direction. 
Introducing vectors $n_\mu$ and $n'_\mu$ such that $n^2=n'^2=0$, $n'n=2$, $v_\mu=p_\mu/M_B=\frac{1}{2}(n_\mu+n'_\mu)$,
one finds that the dominant contribution of charming loops to an FCNC $B$-decay amplitude comes from the
double-collinear configuration \cite{mk2018,m2019,m2022,m2023,wang2022} when the coordinates of the field operators in
$\langle 0|\bar s(y)G_{\mu\nu}(x)b(0)|\bar B_s(p)\rangle$ are alligned along the orthogonal light-cone
directions $x_\mu\sim n_\mu$, $y_\mu\sim n'_\mu$. The 3BS amplitude in the collinear
and the double-collinear kinematics contain the same Lorentz structures \cite{wang2022b} but the
distribution amplitudes corresponding to the collinear and the double-collinear kinematics differ
from each other. 

In this paper we do not consider the double-collinear approximation but make use of the general noncollinear 3BS.
This quantity contains new Lorentz structures and new 3DAs. The $B_s\to\gamma\gamma$ amplitude calculated using the
general noncollinear 3BS differs by terms $O(\lambda_{B_s}/M_B)$ from the amplitude calculated within the double-collinear
approximation.  

\subsection{Collinear 3BS of $B$-meson
\label{Section4A}}
We summarize in this Section well-known results concerning the collinear 3BS that will be used for
constructing a generalization to a noncollinear kinematics appropriate for charming loops in FCNC $B$-decays. 

\subsubsection{The Lorentz structure of the collinear 3BS \label{Section4A1} }
We start with the collinear LC 3BS \cite{braun2017}, where the arguments of the $s$-quark field, $\bar s(y)$, and the gluon field
$G_{\nu\alpha}(x)$ are collinear to each other, $x=u y$, $u\ne 0$ is a number (in this case $x^2=0$ leads to $y^2=0$): 
\begin{eqnarray}
\label{3BScoll}
&&\langle 0| \bar{s}(y) G_{\nu\alpha}(u\, y)\Gamma\, b(0)|\bar{B}_s(p)\rangle=
\frac{f_B M_B^3}{4}
\int D(\omega,\lambda)\, e^{-i\lambda y p -i \omega u y p}
     {\rm Tr}\Bigg\{\gamma_5 \Gamma (1+ \slashed{v})
  \nonumber\\
  &&\hspace{.5cm}
  \times
\bigg[(p_\nu\gamma_\alpha-p_\alpha\gamma_\nu)\frac{1}{M_B}[\Psi_A-\Psi_V] -i\sigma_{\nu\alpha}\Psi_V 
-  \frac{(y_\nu p_\alpha-y_\alpha p_\nu)}{y p} \left(X_A + \frac{\slashed{y}}{y p} M_B W\right) \nonumber\\
&&\hspace{1.0cm}
+ \frac{(y_\nu\gamma_\alpha-y_\alpha\gamma_\nu)}{y p}M_B\left(Y_A + W + \frac{\slashed{y}}{y p} M_B Z\right)
-i\epsilon_{\nu\alpha\mu\beta}\,\frac{y^{\mu}p^{\beta}}{y p}\gamma^5\tilde{X}_A
+i\epsilon_{\nu\alpha\mu\beta}\,\frac{y^{\mu}\gamma^{\beta}}{y p}\gamma^5 M_B\tilde{Y}_A
     \bigg]\Bigg\},
\end{eqnarray}
where $D(\omega,\lambda)$ takes into account rigorous constraints on the variables $\omega$ and $\lambda$: 
\begin{eqnarray}
D(\omega,\lambda)=d\omega\,d\lambda\,\theta(\omega)\theta(\lambda)\theta(1-\omega-\lambda).
\end{eqnarray}
In Eq.~(\ref{3BScoll}), $\Gamma$ is an arbitrary combination of Dirac matrices, $v_\mu=p_\mu/M_B$, and 
all 8 DAs ($\Psi_A$, $\Psi_V$, etc) are functions of two dimensionless arguments $0<\lambda<1$ and $0<\omega<1$.
Here $\lambda$ refers to the momentum carried by the $s$-quark,
and $\omega$ refers to the momentum carried by the gluon. 

The normalization conditions for $\Psi_A$ and $\Psi_V$ have the form \cite{braun2017}:
\begin{eqnarray}
\int D(\omega,\lambda)\, \Psi_A\left(\omega,\lambda\right) = \frac{\lambda_E^2}{3M_B^2},\qquad
\int D(\omega,\lambda)\,  \Psi_V\left(\omega,\lambda\right) = \frac{\lambda_H^2}{3M_B^2}.
\end{eqnarray}
Some of the Lorentz structures in (\ref{3BScoll}) contain factors $x_\mu/xp$ or $x_\mu x_\nu/(xp)^2$.
Since 3BS (\ref{3BScoll}) is a continuous regular function at $x^2=0$ and $xp=0$, the absence
of singularities at $xp\to 0$ leads to the following constraints:  
\begin{eqnarray}
\label{c1}
&&
\int D(\omega,\lambda)\,\big\{X_A,Y_A,\tilde X_A, \tilde Y_A, Z, W\big\}=0,
\nonumber\\
&&
\int D(\omega,\lambda)\, \omega\, \big\{Z, W\big\}=0,\quad \int D(\omega,\lambda)\, \lambda\,\big\{Z, W\big\}=0. 
\end{eqnarray}
These constraints are obtained by expanding the exponential in the integral representation (\ref{3BScoll})
under the condition $x_\mu=u\, y_\mu$ to the necessary order and requiring that the coefficients multiplying terms
singular in $xp\to 0$ vanish.  

\subsubsection{Twist expansion of the 3DAs}
The DAs in~\eqref{3BScoll} have no definite twist. According to~\cite{braun2017}, the distribution amplitudes
might be written as an expansion in functions with definite twist as follows: 
\begin{eqnarray}
  \label{LorentzDAsviatwistDAs}
     \Psi_A(\omega,\lambda) &=& (\phi_3+\phi_4)/2, \nonumber \\    
     \Psi_V(\omega,\lambda) &=& (-\phi_3+\phi_4)/2,\nonumber  \\  
     {X}_A(\omega,\lambda) &=& (-\phi_{3}  - \phi_{4}  + 2\psi_{4} )/2,\nonumber\\
     {Y}_A(\omega,\lambda) &=& (-\phi_{3}  - \phi_{4}  + \psi_{4}  - \psi_{5} )/2,\nonumber \\
     \tilde{X}_A(\omega,\lambda) &=& (-\phi_{3}  + \phi_{4}  - 2\widetilde\psi_{4} )/2,\nonumber \\
     \tilde{Y}_A(\omega,\lambda) &=& (-\phi_{3}  + \phi_{4}  - \widetilde\psi_{4} +\widetilde\psi_{5} )/2,\nonumber \\
     W(\omega,\lambda) &=& (\phi_{4}  - \psi_{4}  - \widetilde\psi_{4}  + \phi_{5}  + \psi_{5}  + \widetilde\psi_{5} )/2,\nonumber \\
\label{twist}    
     {Z}(\omega,\lambda) &=& (-\phi_{3} + \phi_{4} - 2 \widetilde\psi_{4} + \phi_{5} + 2\widetilde{\psi}_{5} - \phi_{6})/4,
\end{eqnarray}
where we keep the contributions up to twist 6 inclusively (the subscript ``$i$'' in $\phi_i$ and $\psi_i$ denote the twist value). 

\subsubsection{Model for DAs entering the collinear 3BS}
The powers of $\omega$ and $\lambda$ determine the behaviour at small quark and gluon momenta.
This power scaling is related to the conformal spins of the fields and remains the key property of the model.

The starting point of our analysis will be the set of DAs in LD model of \cite{braun2017}
for twist 3- and 4, complemented by twist 5 and 6 DAs reconstructed using the constraints (\ref{c1}) \cite{wang2019}:
\begin{eqnarray}
\label{phi3}
\phi_3 &=& \frac{105(\lambda_E^2-\lambda_H^2)}{32\omega_0^7M_B^2}\,
\lambda\omega^2\left(2\omega_0-\omega-\lambda\right)^2\theta\left(2\omega_0-\omega-\lambda\right),\\
  \label{phi4}
  \phi_4 &=& \frac{35(\lambda_E^2+\lambda_H^2)}{32\omega_0^7M_B^2}\,
  \omega^2\left(2\omega_0-\omega-\lambda\right)^3\theta\left(2\omega_0-\omega-\lambda\right),\\
 \label{psi4}
 \psi_4 &=& \frac{35\lambda_E^2}{16\omega_0^7 M_B^2}\,
 \lambda\omega\left(2\omega_0-\omega-\lambda\right)^3\theta\left(2\omega_0-\omega-\lambda\right),\\
\widetilde{\psi}_4 &=& \frac{35\lambda_H^2}{16\omega_0^7 M_B^2}\,
\lambda\omega\left(2\omega_0-\omega-\lambda\right)^3\theta\left(2\omega_0-\omega-\lambda\right),\\
\phi_5 &=& \frac{35\left(\lambda_E^2+\lambda_H^2\right)}{64\omega_0^7 M_B^2}\,
\lambda\left(2\omega_0-\omega-\lambda\right)^{4}\theta\left(2\omega_0-\omega-\lambda\right),\\
\psi_5 &=& - \frac{35\lambda_E^2}{64\omega_0^7 M_B^2}\,
\omega\left(2\omega_0-\omega-\lambda\right)^{4}\theta\left(2\omega_0-\omega-\lambda\right),\\
\widetilde{\psi}_5 &=& -\frac{35\lambda_H^2}{64\omega_0^7 M_B^2}\,
\omega\left(2\omega_0-\omega-\lambda\right)^{4}\theta\left(2\omega_0-\omega-\lambda\right),\\
\label{phi6}
\phi_6 &=& \frac{7\left(\lambda_E^2-\lambda_H^2\right)}{64\omega_0^7 M_B^2}\,
\left(2\omega_0-\omega-\lambda\right)^{5}\theta\left(2\omega_0-\omega-\lambda\right).
\end{eqnarray} 
Dimensionless parameter $\omega_0$ is related to $\lambda_{B}$, the inverse moment of the $B$-meson LC
distribution amplitude, as 
\begin{eqnarray}
  \label{omega0}
\omega_0=\frac{5}{2}\frac{\lambda_{B}}{M_{B}}. 
\end{eqnarray}
For this model, the integration limits take the following form ($2\omega_0<1$): 
\begin{eqnarray}
  \int D(\omega,\lambda)\,\theta\left(2\omega_0-\omega-\lambda\right)\,{(\dots)}=
  \int\limits_0^{2\omega_0} d \omega \int\limits_0^{2\omega_0-\omega}d\lambda\,{(\dots)}. 
\end{eqnarray}
However, as we shall show shortly, certain modifications of the integrated DAs
[which emerge when performing parts integrations
  of the 3BS (\ref{3BScoll})] at large values of $\omega$ and $\lambda$ will be necessary
in order to satisfy the continuity
of the 3BS considered in a noncollinear kinematics. 

%%%%%%%%%%%%%%%%%%%%%%%%%%%%%%%%%%%%%%%%%%%%%%%%%%%%%
%\newpage
%%%                   Section 4.B
%%%
%%%%%%%%%%%%%%%%%%%%%%%%%%%%%%%%%%%%%%%%%%%%%%%%%%%%
\subsection{Generalization to a noncollinear kinematics}
When the coordinates $x$ and $y$ are independent variables, the 3BS has the following decomposition
that involves more Lorentz structures and more 3DAs compared to the collinear approximation:\footnote{We do not include here those structures
that vanish in the collinear limit $x= u y$, such as e.g. $(x_\sigma y_\nu-x_\nu y_\sigma)/xy$. We also do not consider
structures of the type $1/(xp\,yp)$ that may emerge when generalizing the Lorentz structures multiplying
$W$ and $Z$ DAs in (\ref{3BScoll}); according to our analysis the $W$ and $Z$-structures anyway give a marginal
contribution to the FCNC $B$-decay amplitude.} 
\begin{eqnarray}
\label{3BSnoncoll}
&&\langle 0|\bar{s}(y)G_{\nu\alpha}(x)\Gamma\, b(0)|\bar{B}_s(p)\rangle=
\frac{f_B M_B^3}{4}
\int D(\omega,\lambda)\,
e^{-i\lambda y p-i\omega x p}\, {\rm Tr}\Bigg \{\gamma_5\Gamma\,(1 +\slashed{v})
\nonumber \\
&&\times
\bigg[(p_\nu\gamma_\alpha-p_\alpha\gamma_\nu)\frac{1}{M_B}[\Psi_A-\Psi_V] -i\sigma_{\nu\alpha}\Psi_V \nonumber\\
&&-  \frac{(x_\nu p_\alpha-x_\alpha p_\nu)}{x p} \left(X_A^{(x)} + \frac{\slashed{x}}{x p} M_B W^{(x)}\right)
+ \frac{(x_\nu\gamma_\alpha-x_\alpha\gamma_\nu)}{x p}M_B\left(Y^{(x)}_A + W^{(x)} + \frac{\slashed{x}}{x p} M_B Z^{(x)}\right)\nonumber\\
&&-  \frac{(y_\nu p_\alpha-y_\alpha p_\nu)}{y p} \left(X_A^{(y)} + \frac{\slashed{y}}{y p} M_B W^{(y)}\right)
+ \frac{(y_\nu\gamma_\alpha-y_\alpha\gamma_\nu)}{y p}M_B\left(Y^{(y)}_A + W^{(y)} + \frac{\slashed{y}}{y p} M_B Z^{(y)}\right)\nonumber\\
&&-i\epsilon_{\nu\alpha\mu\beta}\,\frac{x^{\mu}p^{\beta}}{x p}\gamma^5\tilde{X}^{(x)}_A
+i\epsilon_{\nu\alpha\mu\beta}\,\frac{x^{\mu}\gamma^{\beta}}{x p}\gamma^5 M_B\tilde{Y}^{(x)}_A
-i\epsilon_{\nu\alpha\mu\beta}\,\frac{x^{\mu}p^{\beta}}{x p}\gamma^5\tilde{X}^{(y)}_A
+i\epsilon_{\nu\alpha\mu\beta}\,\frac{x^{\mu}\gamma^{\beta}}{x p}\gamma^5 M_B\tilde{Y}^{(y)}_A  \bigg]\Bigg\}.
\end{eqnarray}
All invariant amplitudes $\Phi=\Psi_A, \Psi_V,\dots$ are functions of 5 variables, $\Phi(\omega,\lambda,x^2,y^2,xy)$, for
which we may write Taylor expansion in $x^2,y^2,xy$. Here we limit our analysis to zero-order terms in this expansion.
The corresponding zero-order terms in $\Phi$'s are functions of dimensionless arguments
$\lambda$ and $\omega$ and are referred to as the DAs. These DAs contain 
at least the kinematical constraint $\theta(1-\omega-\lambda)$. However, the DAs may have support
in more restricted areas: e.g., the DAs of the LD model Eqs.~(\ref{phi3})--(\ref{phi6}) 
have support in the region $\theta(2\omega_0-\omega-\lambda)$, $2\omega_0<1$.  

Obviously, the functions $\Psi_A$ and $\Psi_V$ in (\ref{3BScoll}) and (\ref{3BSnoncoll}) are the same.
Other DAs in (\ref{3BScoll}) and (\ref{3BSnoncoll}) are related to each other as follows: 
\begin{eqnarray}
  \label{F}
X=X^{(x)}+X^{(y)}, \quad X=\{X_A,Y_A,\tilde X_A,\tilde Y_A\},\quad W=W^{(x)}+W^{(y)}, \quad Z=Z^{(x)}+Z^{(y)}. 
\end{eqnarray}
The amplitude (\ref{3BSnoncoll}) contains two independent kinematical singularities 
$1/xp$ and $1/yp$ in the Lorentz structures. These kinematical singularities of the Lorentz structures
should not be the singularities of the amplitude; this requirement leads to certain constraints which
we are going to consider now. We shall present these constraints for the case when all DAs contain
$\theta(2\omega_0-\omega-\lambda)$, $2\omega_0<1$. 

For the amplitudes of the type $F$, the Lorentz structures of which contain first power of $1/xp$ or $1/yp$,
the appropriate constraint is obtained by expanding the exponential in (\ref{3BSnoncoll}) to zero order
and requiring that the singular terms vanish:
\begin{eqnarray}
%\label{cFx}
\int_0^{2\omega_0-\lambda} d\omega \,X^{(x)}(\omega,\lambda)=0 \qquad \forall \lambda,\nonumber\\
\label{cFy}
\int_0^{2\omega_0-\omega} d\lambda \,X^{(y)}(\omega,\lambda)=0 \qquad \forall \omega.
\end{eqnarray}
Let us introduce the primitives
\begin{eqnarray}
  \label{barFx}
\overline{X}^{(x)}(\omega,\lambda)=\int_0^{\omega} d\omega' {X}^{(x)}(\omega',\lambda),\\
  \label{barFy}
  \overline{X}^{(y)}(\omega,\lambda)=\int_0^{\lambda} d\lambda' {X}^{(y)}(\omega,\lambda'),   
\end{eqnarray}
which, by virtue of (\ref{cFy}), vanish at the boundary of the DA support region: 
\begin{eqnarray}
\label{cPrimitivesF}
\overline{X}^{(x)}(0,\lambda)=\overline{X}^{(x)}(2\omega_0-\lambda,\lambda)=0 \quad \forall \lambda,
\nonumber\\
\overline{X}^{(y)}(\omega,0)=\overline{X}^{(y)}(\omega,2\omega_0-\omega)=0 \quad \forall \omega.  
\end{eqnarray}
For the functions $Z$ and $W$, the Lorentz structure of which contain $1/(xp)^2$ and $1/(yp)^2$,
the exponential should be expanded to first order leading to:
\begin{eqnarray}
  \label{cZ}
  \int\limits_0^{2\omega_0-\lambda} d\omega\, \omega^n Z^{(x)}(\omega,\lambda)=0 \quad \forall \lambda,\quad
  \int\limits_0^{2\omega_0-\omega} d\lambda\, \lambda^n Z^{(y)}(\omega,\lambda)=0\quad \forall \omega,\quad n=0,1.
\end{eqnarray}
By introducing primitives and double primitives  
\begin{eqnarray}
 \overline{Z}^{(x)}(\omega,\lambda)=\int\limits_0^\omega d\omega' Z^{(x)}(\omega',\lambda), \qquad
 \overline{\overline{Z}}^{(x)}(\omega,\lambda)=\int\limits_0^\omega d\omega' \overline{Z}^{(x)}(\omega',\lambda)
 \nonumber\\
 \overline{Z}^{(y)}(\omega,\lambda)=\int\limits_0^\lambda d\lambda' Z^{(y)}(\omega,\lambda'), \qquad
 \overline{\overline{Z}}^{(y)}(\omega,\lambda)=\int\limits_0^\lambda d\omega' \overline{Z}^{(y)}(\omega,\lambda'), 
\end{eqnarray}
one can show that the requirements (\ref{cZ}) lead to the vanishing of these functions at the boundaries of
the DAs support regions: 
\begin{eqnarray}
  \label{cPrimitivesZ}
  \overline{Z}^{(x)}(0,\lambda)=\overline{Z}^{(x)}(2\omega_0-\lambda,\lambda)=0, \qquad
  \overline{\overline{Z}}^{(x)}(0,\lambda)=\overline{\overline{Z}}^{(x)}(2\omega_0-\lambda,\lambda)=0
  \qquad \forall \lambda,\nonumber\\
  \overline{Z}^{(y)}(\omega,0)=\overline{Z}^{(y)}(\omega,2\omega_0-\omega)=0, \qquad
  \overline{\overline{Z}}^{(y)}(\omega,0)=\overline{\overline{Z}}^{(y)}(\omega,2\omega_0-\omega)=0
  \qquad \forall \omega. 
\end{eqnarray}
These relations are verified making use of Eq.~(\ref{cZ}). For instance,  
\begin{eqnarray}
 \overline{\overline{Z}}^{(x)}(2\omega_0-\lambda,\lambda)&=&
  \int\limits_0^{2\omega_0-\lambda}d\omega'\,
  \int\limits_0^{\omega'} d\omega\, Z^{(x)}(\omega,\lambda)\,\theta(\omega<2\omega_0-\lambda)=
  \int\limits_0^{2\omega_0-\lambda} d\omega\, Z^{(x)}(\omega,\lambda)
  \int d\omega'\,\theta(0<\omega<\omega'<2\omega_0-\lambda)\nonumber\\
  &=&
\int\limits_0^{2\omega_0-\lambda} d\omega\, Z^{(x)}(\omega,\lambda)(2\omega_0-\lambda-\omega)=0.  
\end{eqnarray}
The analogous relations are valid for $W$.

For calculating the contribution of (\ref{3BSnoncoll}) to the FCNC amplitude,
the presence of the $1/x p$ and $1/y p$ structures are inconvenient. To facilitate the calculation,
one may perform parts integration in $\omega$ for the structure containing $1/xp$ and in $\lambda$
for the structure containing $1/yp$. The conditions (\ref{cPrimitivesF}) and (\ref{cPrimitivesZ})
lead to the absence
of the surface terms when performing the parts integrations.  
So, we can rewrite (\ref{3BSnoncoll}) in the convenient form not containing the $1/xp$ and $1/yp$ factors:
\begin{eqnarray}
\label{3BSnoncoll2}
&&\langle | \bar{s}(y) G_{\nu\alpha}(x)\Gamma\,b(0) |\bar{B}_s(p)\rangle=
\frac{f_B M_B^3}{4}
\int D(\omega,\lambda)\; e^{-i\lambda\, (y p) -i \omega\, (x p)}
{\rm Tr}\Bigg\{\gamma_5\Gamma\,(1+\slashed{v})\bigg[
\nonumber\\
&&\times (p_\nu\gamma_\alpha-p_\alpha\gamma_\nu)\frac{1}{M_B}[\Psi_A-\Psi_V] 
-i\sigma_{\nu\alpha}\Psi_V
\nonumber\\
 && - i (x_\nu p_\alpha-x_\alpha p_\nu) \left(\overline{X}_A^{(x)} + i\slashed{x} M_B\overline{\overline{W}}^{(x)}\right) 
        + i (x_\nu\gamma_\alpha-x_\alpha\gamma_\nu)M_B\left( \overline{Y}_A^{(x)} + \overline{W}^{(x)} +
        i \slashed{x} M_B\overline{\overline{Z}}^{(x)}\right)
\nonumber\\ 
&&      - i (y_\nu p_\alpha-y_\alpha p_\nu) \left(\overline{X}_A^{(y)} + i\slashed{y} M_B\overline{\overline{W}}^{(y)}\right) 
+ i (y_\nu\gamma_\alpha-y_\alpha\gamma_\nu)M_B\left( \overline{Y}_A^{(y)} + \overline{W}^{(y)} + i \slashed{y}M_B\overline{\overline{Z}}^{(y)}\right)
\nonumber\\ 
&&     +\epsilon_{\nu\alpha\mu\beta}\,x^{\mu}p^{\beta}\gamma^5\overline{\tilde{X}}_A^{(x)}
      -\epsilon_{\nu\alpha\mu\beta}\,x^{\mu}\gamma^{\beta}\gamma^5 M_B\overline{\tilde{Y}}_A^{(x)}
      +\epsilon_{\nu\alpha\mu\beta}\,y^{\mu}p^{\beta}\gamma^5\overline{\tilde{X}}_A^{(y)}
      -\epsilon_{\nu\alpha\mu\beta}\,y^{\mu}\gamma^{\beta}\gamma^5M_B\overline{\tilde{Y}}_A^{(y)}
\bigg]\Bigg\}.
\end{eqnarray}
We emphasize that if the conditions (\ref{cFy}) and (\ref{cZ}) are not fulfilled, 
then the parametrizations (\ref{3BSnoncoll}) and (\ref{3BSnoncoll2}) are not equivalent:
they differ by a nonzero surface term. 
Noteworthy, the requirements of the absence of singularities at $xp\to 0$ and $yp\to 0$ and the continuity
of the 3BS (\ref{3BSnoncoll}) at $x^2=0$, $y^2=0$, $xp=0$ and $yp=0$ lead to a number of constraints on the DAs
which guarantee the equivalence of the forms (\ref{3BSnoncoll}) and (\ref{3BSnoncoll2}). 
 
\subsection{Adapting the LD model for the case of noncollinear 3BS \label{Section4C}}
First, let us point out that the primitives calculated for the DAs of the LD model (\ref{phi3})--(\ref{phi6}) 
do not satisfy the constraints (\ref{cPrimitivesF}) and (\ref{cPrimitivesZ}).
As already emphasized, the parametrizations (\ref{3BSnoncoll}) and (\ref{3BSnoncoll2}) are then
not equivalent and cannot be both correct. So, there are two different possibilities to handle this situation:

\subsubsection{Scenario I} 
Since the behaviour of the DAs of the definite twist at small $\omega$ and $\lambda$ are fixed,
we choose the following procedure:\\
\noindent 1. The functions $\Psi_A(\omega,\lambda)$ and $\Psi_V(\omega,\lambda)$ remain intact.\\
\noindent 2. We split each function $\Phi$ $(\Phi=X_A,Y_A,\tilde X_A,\tilde Y_A, Z,W)$ into
$\Phi^{(x)}(\omega,\lambda)$ and $ \Phi^{(y)}(\omega,\lambda)$ as follows:\footnote{In principle,
one can take different $w_x$ for each of the 3DAs. We will not discuss this possibility but will assume that
$w_x$ is the same for all 3DAs.}
\begin{eqnarray}
\Phi^{(x)}(\omega,\lambda)= w_x \Phi(\omega,\lambda),\quad
\Phi^{(y)}(\omega,\lambda)= w_y \Phi(\omega,\lambda), \quad w_x+w_y=1.
\end{eqnarray}
\noindent 3. We make use of the DAs of LD model for the functions $\Phi(\omega,\lambda)$
and calculate primitives of higher orders. Taking into account that higher primitives 
obtained in this way do not satisfy the constraints (\ref{cPrimitivesF}) and (\ref{cPrimitivesZ}),
we modify them in the way explained below and allow them to depend on one additional parameter $a$. 

\noindent 4. We still want that after taking the appropriate derivatives of the modified primitives,
we reproduce Eq.~(\ref{3BSnoncoll}) with 6 modified functions $(X_A,Y_A,\tilde X_A,\tilde Y_A, Z,W)$
which in turn also depend on the additional parameter $a$. The simplest consistent scheme that fulfills this requirement is the following: 

\noindent 
$\bullet$
For the functions in Eq.~(\ref{3BSnoncoll}) containing factors $1/xp$ and $1/yp$ (i.e.
$X=X_A, Y_A, \widetilde{X}_A, \widetilde{Y}_A$) the primitives entering
Eq.~(\ref{3BSnoncoll2}) are constructed as follows: 
\begin{eqnarray}
  &\overline{X}^{(x)} = \frac{\partial}{\partial \lambda}
  \bigg[R(\omega,\lambda, a)\int\limits_0^\omega d\omega' \int\limits_0^\lambda  d\lambda' X(\omega',\lambda')\bigg],
  \nonumber\\ 
  & \overline{X}^{(y)} = \frac{\partial}{\partial \omega}\bigg[ R(\omega,\lambda, a)
    \int\limits_0^\omega d\omega' \int\limits_0^\lambda d\lambda' X(\omega',\lambda')\bigg].
\end{eqnarray}
\noindent 
$\bullet$ For the functions in Eq.~(\ref{3BSnoncoll}) containing
factors $1/xp$ and $1/yp$ and $1/(xp)^2$ and $1/(yp)^2$
(i.e., $Z$ and $W$) the primitives entering Eq.~(\ref{3BSnoncoll2}) are constructed in a different way
(the same formulas for $W$):
\begin{eqnarray}
  \overline{\overline{Z}}^{(x)} &=&
  \frac{\partial^2}{\partial \lambda^2} \bigg[R(\omega,\lambda, a)
    \int\limits_0^\omega d\omega' \int\limits_0^\lambda  d\lambda'
    \int\limits_0^{\omega'} d\omega'' \int\limits_0^{\lambda'} d\lambda'' Z(\omega'',\lambda'')\bigg],
 \nonumber \\
  \overline{\overline{Z}}^{(y)} &=& \frac{\partial^2}{\partial \omega^2}\bigg[ R(\omega,\lambda, a)
    \int\limits_0^\omega d\omega'\int\limits_0^\lambda  d\lambda' \int\limits_0^{\omega'} d\omega''\int\limits_0^{\lambda'} 
    d\lambda'' Z(\omega'',\lambda'')\bigg],
 \nonumber   \\
  \overline{Z}^{(x)} &=& \frac{\partial}{\partial \omega}\frac{\partial^2}{\partial \lambda^2}\bigg[R(\omega,\lambda, a)
    \int\limits_0^\omega  d\omega'\int\limits_0^\lambda d\lambda' \int\limits_0^{\omega'}  d\omega'' \int\limits_0^{\lambda'}
    d\lambda'' Z(\omega'',\lambda'')\bigg],
  \nonumber  \\
  \overline{Z}^{(y)} &=& \frac{\partial}{\partial \lambda}\frac{\partial^2}{\partial \omega^2}\bigg[ R(\omega,\lambda, a)
    \int\limits_0^\omega  d\omega'\int\limits_0^\lambda d\lambda' \int\limits_0^{\omega'}  d\omega''\int\limits_0^{\lambda'}
    d\lambda'' Z(\omega'',\lambda'')\bigg].
\end{eqnarray}
The function $R(\omega,\lambda,a)$ is chosen in the form   
\begin{eqnarray}
  \label{R}
R(\omega,\lambda,a)=\frac{2}{1+\exp\left(\frac{a}{2\omega_0-\omega-\lambda}\right)},  
\end{eqnarray}
such that for $a\ne 0$ the function itself and all its derivatives vanish at the boundary $\omega+\lambda=2\omega_0$. 
Respectively, for $a\ne 0$, all modified higher primitives of the necessary order
vanish at the boundary $2\omega_0-\omega-\lambda=0$,
providing the absence of the $1/xp$ and $1/yp$ singularities at $xp\to 0$ and $yp\to 0$ and
the continuity of the 3BS at the point $x^2=0$, $y^2=0$, $xp=0$ and $yp=0$. 

For small values of the parameter $a$, our model DAs reproduce well the collinear DAs including their magnitudes 
and power behaviour at {\it small} $\omega$ and $\lambda$ but (strongly) deviate from them near the upper boundary.

\subsubsection{Scenario II}
One declares Eq.~(\ref{3BSnoncoll2}) as the starting point but calculates the necessary primitives
using the 3DAs from Eq.~(\ref{3BSnoncoll}). Then, however, Eq.~(\ref{3BSnoncoll}) itself is not complete
and should contain nonzero surface terms obtained by rewriting Eq.~(\ref{3BSnoncoll2}) to the form (\ref{3BSnoncoll}). 
These surface terms guarantee the absence of singularities in (\ref{3BSnoncoll}) at $xp\to 0$ and $yp\to 0$.
This scenario does not seem to us logically satisfactory: For instance, the double-collinear limit
\cite{wang2022,m2023} that may be readily taken in the 3BS given by Eq.~(\ref{3BSnoncoll}) is not reproduced
by  Eq.~(\ref{3BSnoncoll2}) if the surface terms are nonzero!
Nevertheless, for comparison we also present the results for the
form factors calculated using this prescription referred to as Scenario II.
It should be emphasized that for one and the same set of 3DAs,
the form factors obtained using Scenario I in the limit $a\to 0$ do not reproduce the form factors obtained
using Scenario II: the difference amounts to certain surface terms that arise in the limit $a\to 0$.

%\newpage
%%%%%%%%%%%%%%%%%%%%%%%%%%%%%%%%%%%%%%%%%%%%%%%%%%%%%%
%  SECTION 5
%%%%%%%%%%%%%%%%%%%%%%%%%%%%%%%%%%%%%%%%%%%%%%%%%%%%%%%%%%%%%%%%%%%%%%%%%%
\section{\label{sec:5}Results for the $B_s\to \gamma\gamma$ amplitude and $\delta C_{7\gamma}$}
We are ready to evaluate the form factors $H_{A,V}$. We shall also calculate
the penguin form factor $F_{T}$, and in the end, the charming-loop correction to $C_{7\gamma}$. 
\subsection{The charming-loop form factors $H_{A,V}(0,0)$}
Using Eqs.~(\ref{Acharm3a}) and (\ref{Acharm3b}) and calculating the trace in (\ref{3BSnoncoll2}) for
\begin{eqnarray}
  \Gamma=\gamma^\eta(\slashed k+m_s)\gamma^\mu(1-\gamma_5), 
\end{eqnarray}
leads to the $B_s\to \gamma\gamma$ amplitude expressed via the DAs and their primitives. 

The expression (\ref{3BSnoncoll2}) contains powers of $x$ and/or $y$, but these are easy to handle.
Let us go back to Eq.~(\ref{Acharm3b}): Any factor $x_\alpha$ may be represented as
$x_\alpha\to \frac{\partial}{\partial \kappa_\alpha}e^{-i\kappa x}$
then taking the $\kappa$-derivative over to $\overline{\Gamma}_{cc}^{\mu\nu\rho\alpha}(\kappa,q)$ by parts integration in $\kappa$.
Any factor $y_\alpha$ may be represented as $y_\alpha\to \frac{\partial}{\partial k_\alpha}e^{-i k y}$, then 
moving the derivative onto the strange-quark propagator by parts integration in $k$.
Doing so, we get rid of all powers of $x$ and $y$, and 
the integrals over $x$ and $y$ in (\ref{Acharm3b}) then lead to the $\delta$ functions:
\begin{eqnarray}
\int dx \rightarrow (2\pi)^4\delta\left(\kappa  + \omega p\right),\qquad 
\int dy \rightarrow (2\pi)^4\delta\left(k - q' + \lambda p\right), 
\end{eqnarray}
which allow us to further take the integrals over $\kappa$ and $k$.
In the end, the invariant functions $H_i$, $i={A,V}$ are given by integral
representations of the form 
\begin{eqnarray}
  \label{Hi}
  H_i(q^2,q'^2) =
  \int\limits_0^{2\omega_0} d\omega\int\limits_0^{2\omega_0-\omega} d\lambda\int\limits_0^1
  d\xi \int\limits_0^{1-\xi}d\eta\;h_i(\omega,\lambda,\xi,\eta\,|\,q^2,q'^2)
\end{eqnarray}
where $h_i$ are linear combinations of the DAs and their primitives entering Eq.~(\ref{3BSnoncoll2}). 
Eq.~(\ref{Hi}) gives now the form factors $H_{A,V}$ as the convolution
integrals of the known expression for the charming loops and the DAs and its primitives
from (\ref{3BSnoncoll2}). We make use of the modified LD model described above.

%%%%%%%%%%%%%%%%%%%%%%%%%%%%%%%%%%
%%%  Plot FFs
%%%%%%%%%%%%%%%%%%%%%%%%%%%%%%%%%%%
\begin{figure}[b!]
\begin{center}
\begin{tabular}{cc}   
  \includegraphics[height=5cm]{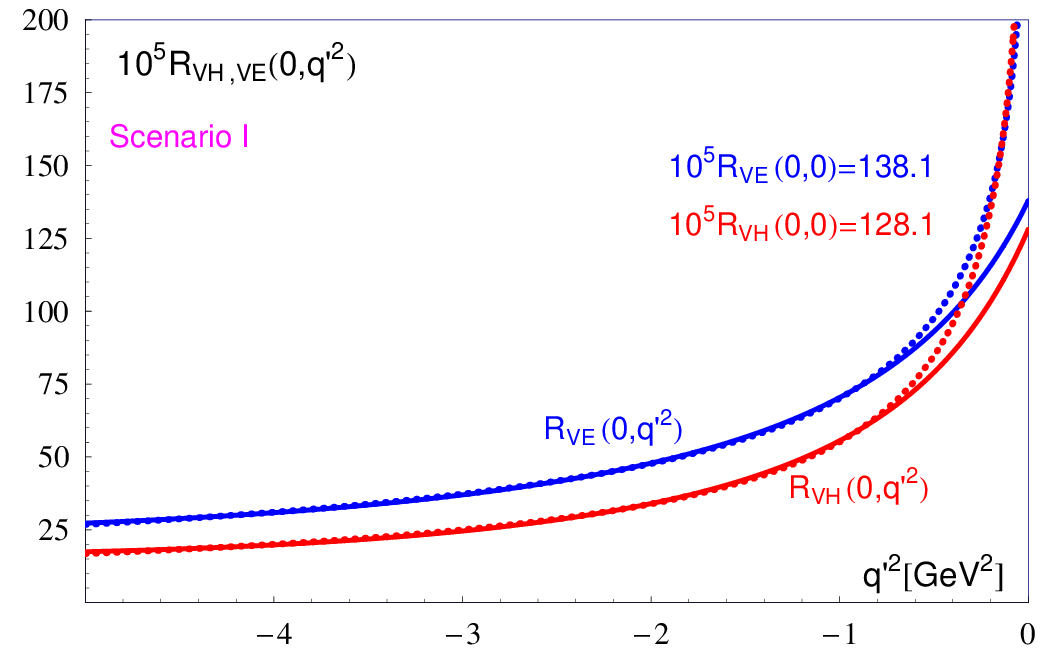}           &  \includegraphics[height=5cm]{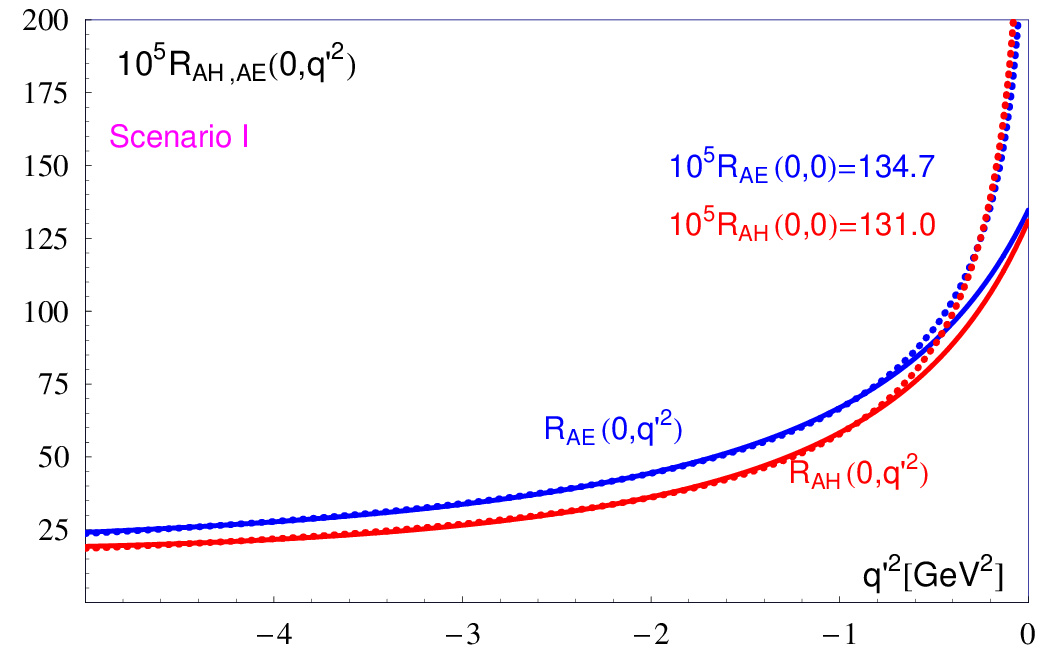}\\
  (a)  &   (b)  \\
  \includegraphics[height=5cm]{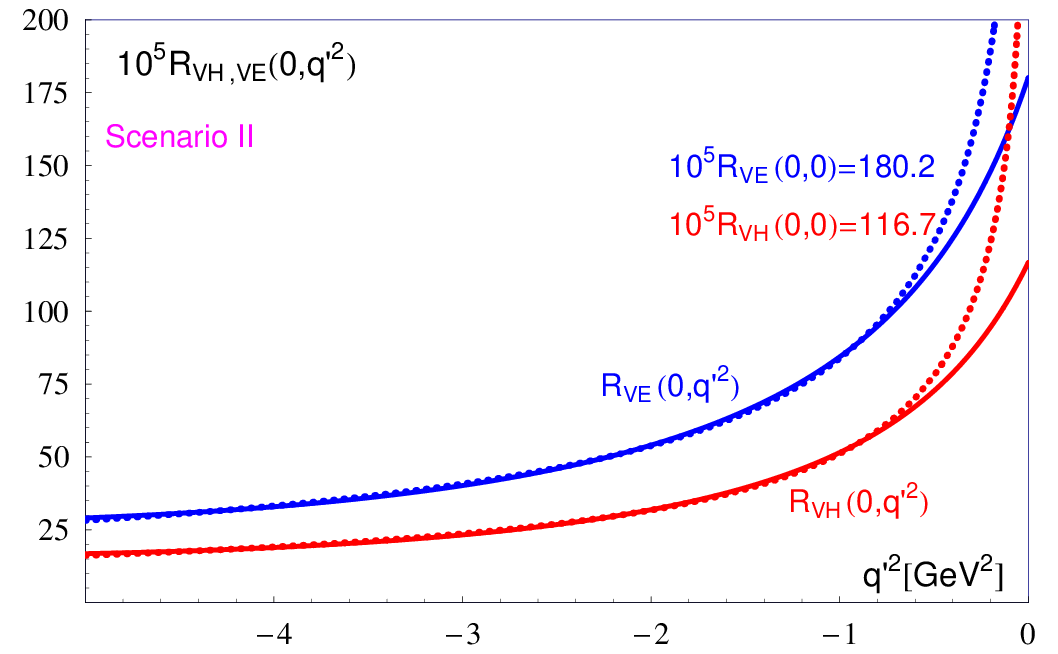}     &  \includegraphics[height=5cm]{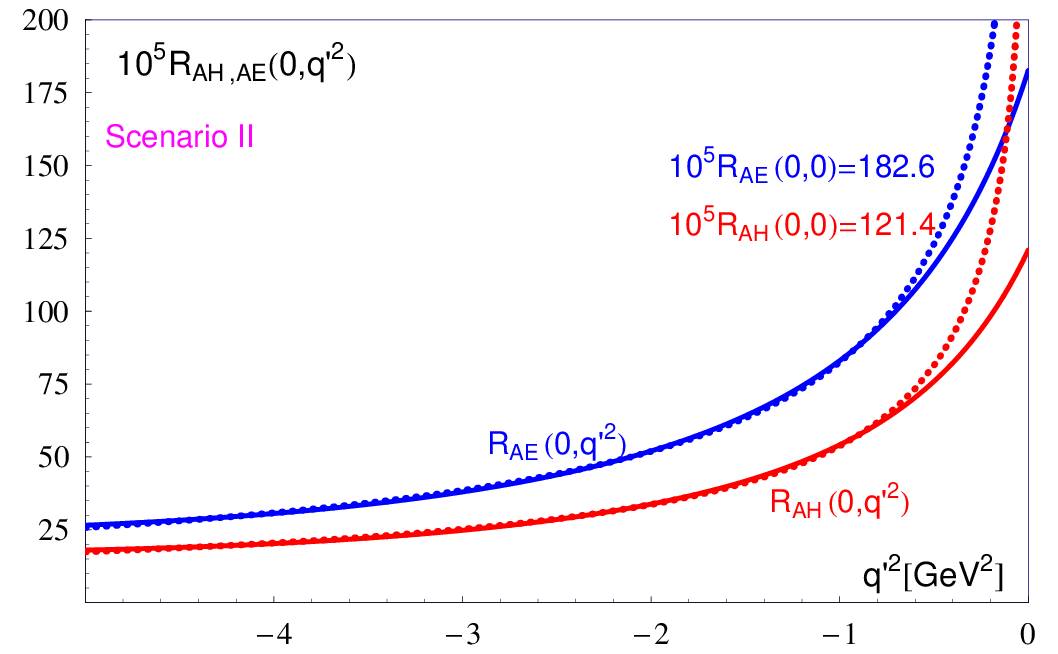}\\
  (c)  &   (d)  \\
  \includegraphics[height=5cm]{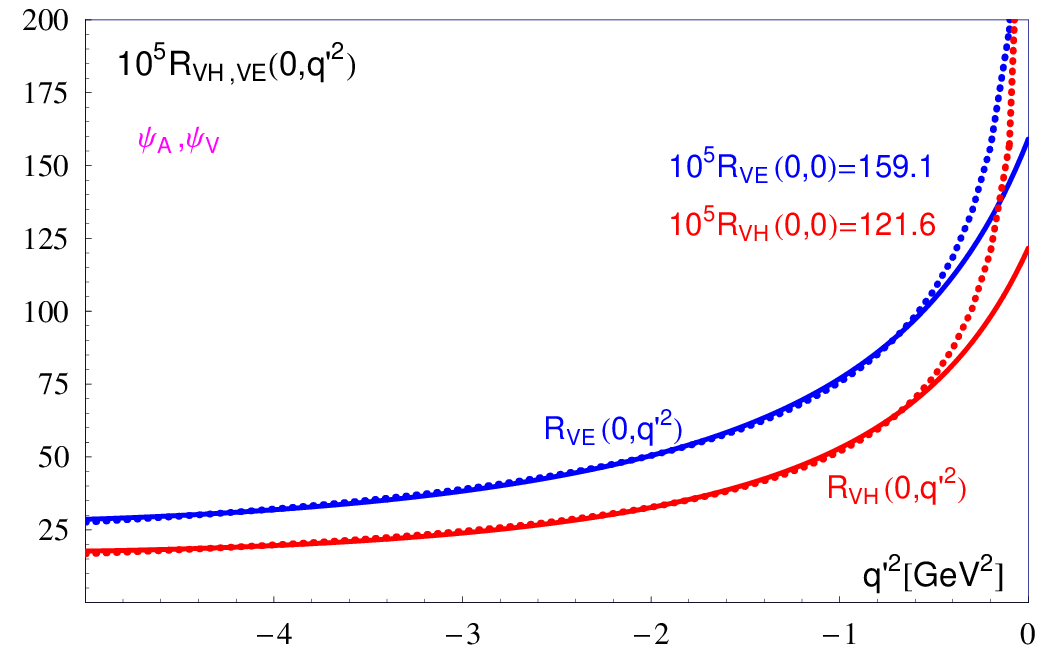} &  \includegraphics[height=5cm]{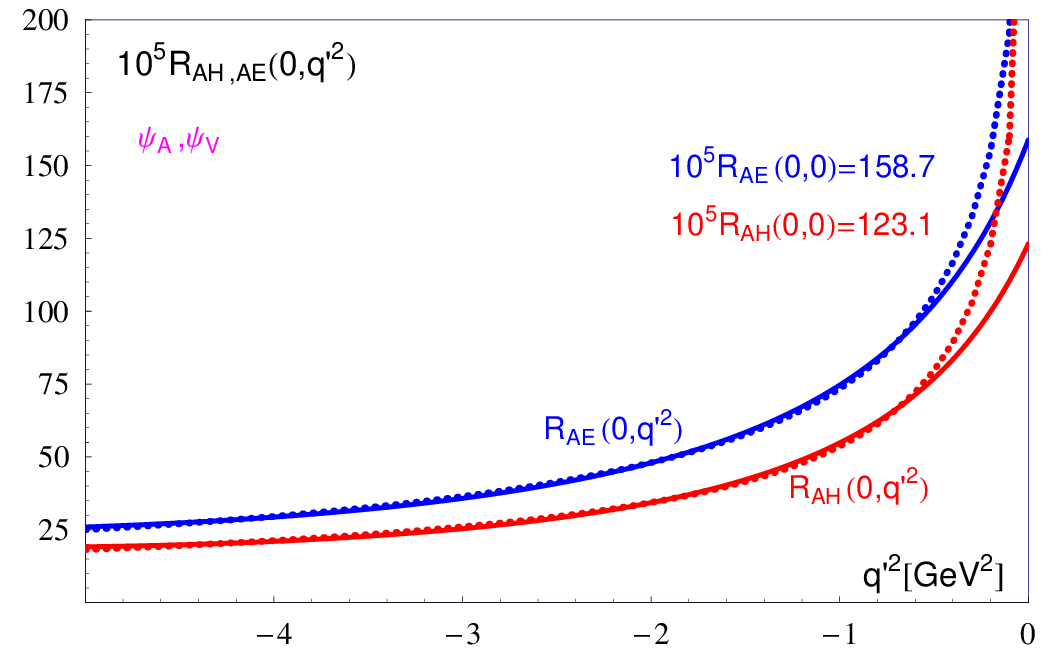}\\
  (e)  &   (f)  
\end{tabular}
\caption{\label{Plots:H} The contributions $R_{iE}$ and $R_{iH}$ to the form factors $H_i$ [$i=A,V$]
  defined according to Eq.~(\ref{Ri}) for different scenarios of treating the $B$-meson 3DA.
  (a,b): The appropriate modifications of the 3DAs $X_A$, $Y_A$, {\it etc.} are taken into account (Scenario I). 
  (c,d): The form factors are calculated using Eq.~(\ref{3BSnoncoll2}) with the primitives obtained from
  $X_A$, $Y_A$, {\it etc.} without modifications (Scenario II).
(e,f): Only the contributions of $\Psi_A$ and $\Psi_V$ are taken into account. 
  Dashed lines show the calculation results for $\lambda_{B_s}=0.45$ GeV, $a=0.05$, $w_x=w_y=0.5$ and
  other inputs from Table \ref{Table:Parameters}, and solid lines are the fits using Eq.~\eqref{fit}.
\label{Plot:H}}
\end{center}
\end{figure}
Obviously, the form factors $H_{A}$ and $H_V$ depend linearly on $\lambda_E^2$ and $\lambda_H^2$. So, we present the
results for the form factors $R_{iE}$ and $R_{iH}$ defined as follows:
\begin{eqnarray}
  \label{Ri}
  H_{i}(0,q'^2)&=&R_{iE}(0,q'^2)\lambda_E^2+R_{iH}(0,q'^2)\lambda_H^2, \qquad i=A,V. 
\end{eqnarray}  
The form factors $R_{iE,iH}$ depend on $\lambda_{B_s}$, but do not contain dependence on $\lambda_H$ and $\lambda_E$.
These three parameters, however, are expected to be strongly correlated and we take into account this
correlation later when evaluating $\delta C_{7\gamma}$.

All numerical inputs necessary for calculating $R_i$ are summarized in Table \ref{Table:Parameters}.
\begin{table}[hb!]
\centering 
\caption{Input parameters for the calculation of the form factors $R_{iE,iH}$ defined in Eq.~(\ref{Ri}).
  \label{Table:Parameters}}
\begin{tabular}{|c|c|c|c|c|c|c|c|}
  \hline
$\overline{m}_c(\overline{m}_c)$ & $\overline{m}_s(2\,{\rm GeV})$&$M_{B_s}$&$M_{\phi}$&$f_{B_s}$&$\lambda_{B_s}(1\,{\rm GeV})$& $a$  &  $w_x$ \\
  $1.30$ GeV    &   0.1 GeV    & 5.3 GeV  & 1.020 GeV &  0.23 GeV & $0.45\pm 0.15$ GeV  &  $0.05\pm 0.05$ & $0.5\pm 0.5$  \\
  \hline
\end{tabular}
\end{table}
Making use of the estimate for the ratio  $\lambda_{B_s}/\lambda_B=1.19\pm 0.04$ \cite{khodjamirian2020} and
available theoretical results for $\lambda_B$ \cite{kou,braun2004,beneke2011,zwicky2021,im2022}, 
we consider the range $\lambda_{B_s}=0.45\pm 0.15$ GeV with $\lambda^0_{B_s}=0.45$ GeV as the benchmark value. 

Presently, the theoretical knowledge of noncollinear 3DAs is poor to give good arguments for the choice of the
function $R$, Eq.~(\ref{R}), and the weight factors $w_x$. We present our results for the form factors
for $w_x=0.5$ and reflect the sensitivity to its variations in the ranges $0<w_x<1$ in the final uncertainties. 
We restrict the parameter $a$ by the requirement that the 3DAs are affected by, say, not more than 10\% in the
region of ``small'' $\lambda$ and $\omega$, e.g., at $\lambda/(2\omega_0), \omega/(2\omega_0)\le 0.1$.
This leads to the restriction $a\le 0.1$. So we set the benchmark point $a=0.05$ and allow its
variations in the range $0 \le  a \le 0.1$. We shall see that the sensitivity of the calculated form factors
to $a>0.1$ is rather mild. 

Other parameters in Table~\ref{Table:Parameters} are taken from \cite{pdg} and \cite{lattice}.

We are interested in obtaining the form factors $R_{iE,iH}$ at $q^2=0$ and $q'^2=0$.
Whereas $q^2=0$ may be readily set in the integral representation (\ref{Hi}), this integral representation
is not expected to give reliable predictions at $q'^2=0$:
The $s$-quark propagator at $q'^2=0$ is not sufficiently far off-shell and at $q'^2\to 0$ one observes
a steep rise of $R_{iE,iH}$ related to the nearby quark singularity.
This rise is unphysical as the nearest physical singularity in the form factors is located at $q'^2=M_\phi^2$. 
To avoid this problem, we choose the following strategy: we take the results of our calculation for 
$R_{iH,iE}(0,q'^2)$ in the interval $-5< q'^2({\rm GeV}^2) <-0.7$ and extrapolate them numerically to $q'^2=0$
making use of a modified pole formula which takes into account the presence of the physical pole at $q'^2=M_\phi^2$:
\begin{equation}
  \label{fit}
  F(q^2) = \frac{F(0)}{\left(1 - q^2/M_R^2\right)\left(1 - \sigma_1(q^2/M_R^2) - \sigma_2(q^2/M_R^2)^2\right)}. 
\end{equation}
In this way we obtain $R_{iE,iH}(0,0)$.

Figure~\ref{Plot:H} shows the results of our direct calculation 
and the fits obtained using Eq.~(\ref{fit}). Figure~\ref{Plot:H}(a,b) gives the
form factors evaluated with the modified
3DAs for $a=0.05$ (Scenario I).
Figure~\ref{Plot:H}(c,d) shows the results for unmodified functions (Scenario II). For comparison,
Fig.~\ref{Plot:H}(e,f) shows the contribution of the 3DAs $\psi_A$ and $\psi_V$. Good news is that the latter give the 
dominant contribution (which does not depend on $w_x$ and $a$ and thus reduces the uncertainty in the $R_i$ related to
the values of these parameters). The contribution of other Lorentz 3DAs is however not negligible and depend on the
way one handles these 3DAs: e.g. for $R_{AE}(0,0)$ in Scenario I they give a negative correction of 30\%.  
Appendix \ref{Appendix_Anathomy} presents details of the contributions of different Lorentz 3DAs to the
form factors $R_{iE,iH}(0,q'^2)$ at $q'^2=-2$ GeV$^2$ for Scenarios I and II. 

\begin{figure}[ht]
\begin{center} 
  \includegraphics[height=5cm]{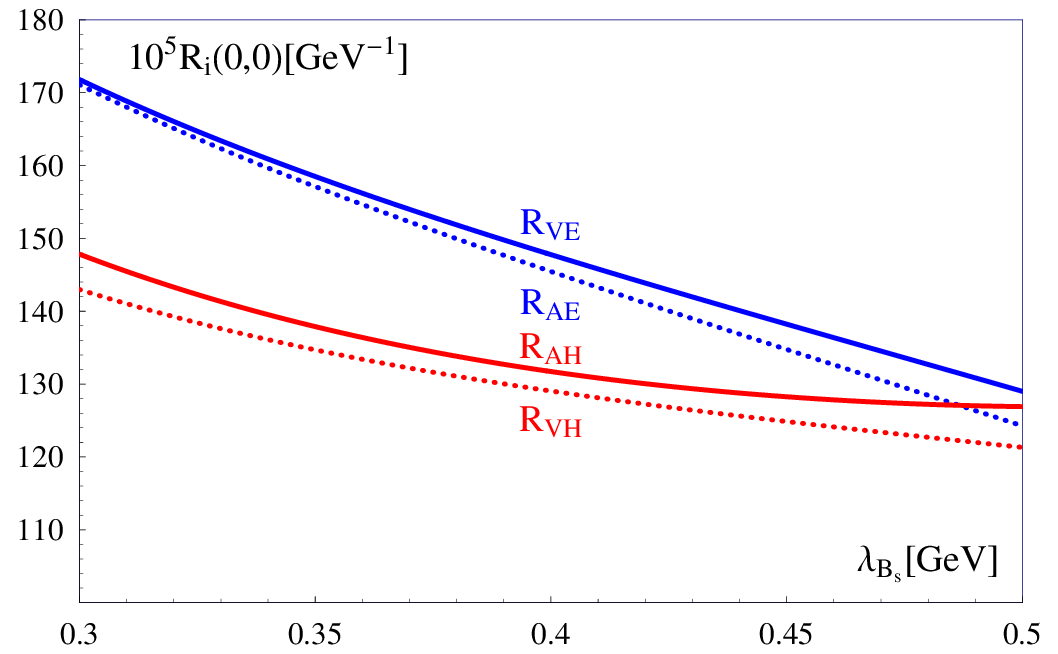}\qquad \includegraphics[height=5cm]{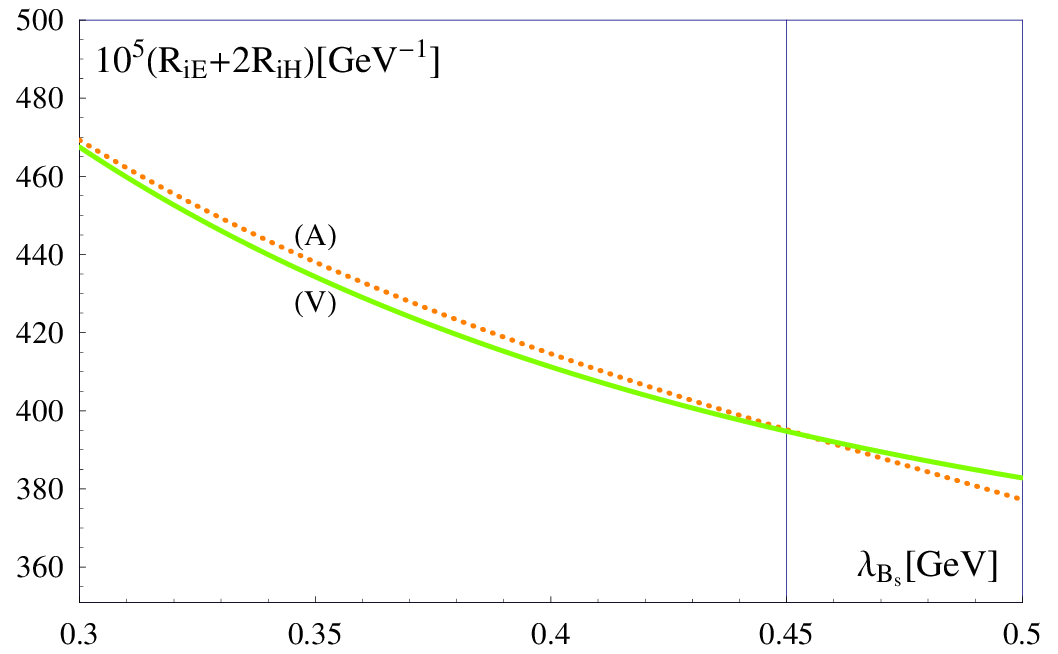} 
  \caption{\label{Plot:RvslambdaBs}
The dependence on parameter $\lambda_{B_s}$ of the form factors obtained in Scenario I: 
  (a) $R_{iE,iH}(0,0)$, $i=A,V$;
  (b) Linear combination $R_{iE}(0,0)+2R_{iH}(0,0)$ that determines $\delta C_{7\gamma}$ taking into account
  approximate relation $\lambda_H^2\simeq 2\lambda_E^2$.}
\end{center}
\end{figure}

Fig.~\ref{Plot:RvslambdaBs} demonstrates the sensitivity of $R_{iE,iH}(0,0)$ to $\lambda_{B_s}$. 
Since the form factors at $q'^2=0$ are obtained via extrapolation, we take $\lambda_{B_s}$ from
the range $0.3<\lambda_{B_s}({\rm GeV})<0.5$, calculate $R_{iE,iH}(0,q'^2)$ in the range $-5 <q'^2({\rm GeV}^2)<-0.7$
and then extrapolate to $q'^2=0$ for each value of $\lambda_{B_s}$. 

\begin{figure}[ht]
\begin{center} 
 \includegraphics[height=5cm]{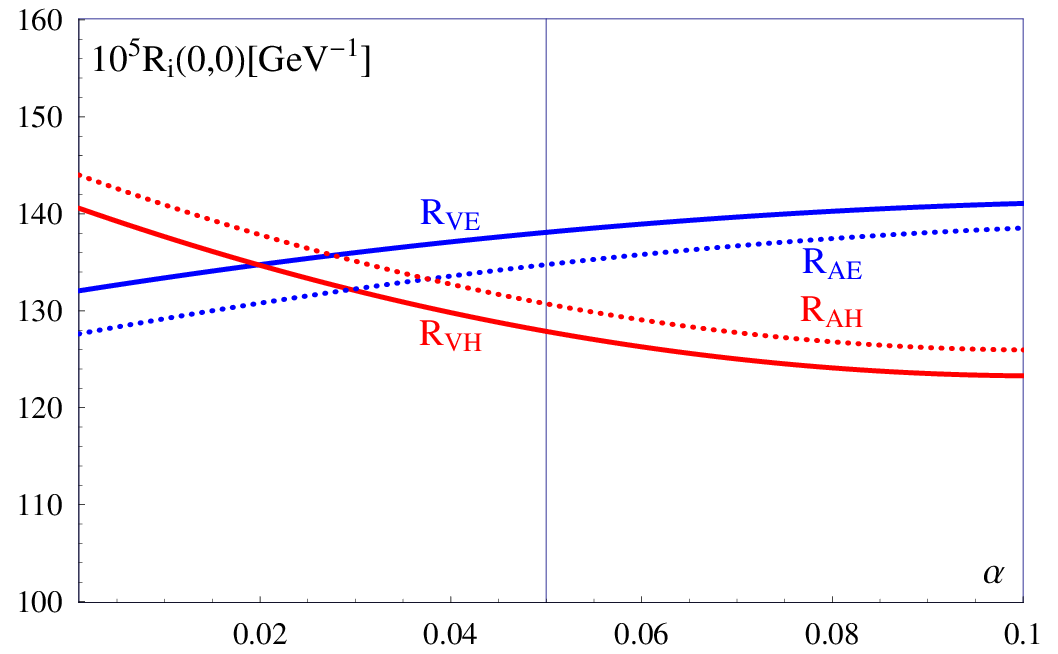} \qquad \includegraphics[height=5cm]{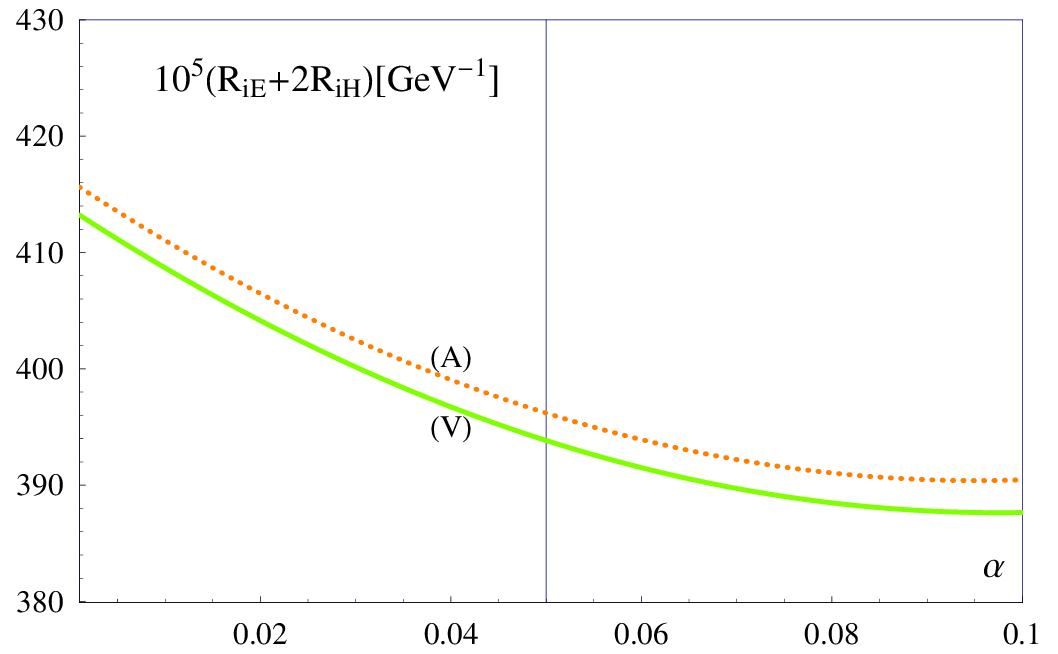} 
\caption{\label{Plot:Rvsa}
  The dependence on parameter $a$ of the form factors obtained in Scenario I: 
  (a) $R_{iE,iH}(0,0)$, $i=A,V$;
  (b) Linear combination $R_{iE}(0,0)+2R_{iH}(0,0)$ that determines $\delta C_{7\gamma}$ taking into account
  approximate relation $\lambda_H^2\simeq 2\lambda_E^2$.}
\end{center}
\end{figure}

The same procedure applies to the dependence on parameter $a$ [enters the function $R$, Eq.~(\ref{R})]
shown in Fig.~\ref{Plot:Rvsa}. We restirct $a$ in the range $0\le a\le 0.1$, but, as seen from
Fig.~\ref{Plot:Rvsa}, the sensitivity to the value of $a$ in the region $a>0.01$ is rather weak anyway. 

The uncertainties in our theoretical predictions for the form factors thus come from the following sources:
\begin{itemize}
\item[(i)] The sensitivity to the precise way one handles the 3BS at large values of $\omega$ and $\lambda$.
This is probed by comparing the results obtained with our benchmark  Scenario I with those from Scenario II.
\item[(ii)]
  The sensitivity to the precise functional form of the $B$-meson 3DAs; this is probed by using as
  our benchmark model (\ref{phi3})-(\ref{phi6})
[Model IIB from \cite{braun2017}] and comparing with a different model [Model IIA given by Eq.~(5.23) from \cite{braun2017}].
The uncertaintes related to (i) and (ii) are denotes as [3DA]. 
\item[(iii)]
  The sensitivity to the numerical values of the parameters of the 3DAs, mainly parameter $a$ of the function $R$, Eq.~(\ref{R}).
  Notice that we do not perform any averaging over $\lambda_{B_s}$; we keep the full dependence on this parameter in the form factors
  and in our final result for the correction $\delta C_{7\gamma}$. 
\item[(iv)]
The sensitivity to the extrapolation procedure from the interval where the form factors may be calculated by our 
approach to the physical point $q'^2=0$. We make use of the calculations of the form factors in the interval
$-5\le q'^2({\rm GeV}^2)\le -0.7$. However, the value at $q'^2=0$ obtained by the
extrapolation is sensitive to the precise choice of the upper boundary of this interval. 
For instance, moving the upper boundary in the range $(-1\div -0.5)$ GeV$^2$ leads to
the variations of the extrapolated value of the form factor at $q'^2=0$ by $\pm 5\%$.
We therefore assign the additional uncertainty of $5\%$ (denoted as [extr]) related to the extrapolation procedure. 
\end{itemize}
Table \ref{Table:Results} compares the form factors at $q^2=q'^2=0$ for two different sets of 3DAs and
for two different Scenarios to treat the 3DAs. 
\begin{table}[hb!]
\centering 
\caption{Our results for the form factors $R_{iE,iH}$ in GeV$^{-1}$ defined in Eq.~(\ref{Ri}).
  In addition to the results obtained in Scenario I and Scenario II for the
  basic set of 3DAs given in (\ref{phi3})--(\ref{phi6}), referred to as (B),
  we present the results obtained using Scenario I for 3DAs of Model IIA of Eq.~(5.23) from \cite{braun2017},
  referred to as (A). 
  \label{Table:Results}}
\begin{tabular}{|c|c|c|c|c|c|c|c|}
  \hline
  & $10^5 R_{VE}$ & $10^5 R_{VH}$ & $10^5(R_{VE}+2R_{VH})$ & $10^5 R_{AE}$ & $10^5 R_{AH}$ & $10^5(R_{AE}+2R_{AH})$ \\
  \hline
Scenario I,  3DA (B)  &      138.1   &  128.1      &          394.3       &    134.7     & 131.0        &   396.7   \\  
Scenario I,  3DA (A)  &      146.4   &  133.7      &          413.6       &    142.7     & 137.0        &   425.4   \\
Scenario II, 3DA (B) &      180.2   &  116.7      &          413.8       &    182.6     & 121.4        &   416.7   \\
  \hline
\end{tabular}
\end{table}
The individual form factors $R_{iE}$ and $R_{iH}$ demonstrate a sizeabe dependence on the Scenario. However,
in the combinations appropriate for the calculation of $\delta C_{7\gamma}$, $R_{VE}+2R_{VH}$ and $R_{AE}+2R_{AH}$
these uncertainties cancel to great extent and we obtain for the benchmark value $\lambda_{B_s}=0.45$ GeV: 
\begin{eqnarray}
\label{RVE+2RVH}
  10^5(R_{VE}+2R_{VH})\,[{\rm GeV}^{-1}]&=&410\pm 10\,[{\rm 3DA}]\pm 15\,[a]\pm 20\,[{\rm extr}]\to 410\pm 30,  
  \\
  \label{RAE+2RAH}
  10^5(R_{AE}+2R_{AH})\,[{\rm GeV}^{-1}]&=&415\pm 15\,[{\rm 3DA}]\pm 15\,[a]\pm 20\,[{\rm extr}]\to 415\pm 30,
\end{eqnarray}
and we have slightly increased our final ``theoretical uncertainty''. In the end, for a given value of
the $\lambda_{B_s}$, we predict the combination of the form factors (\ref{RVE+2RVH}) and (\ref{RAE+2RAH})
relevant for the calculation of $C_{7\gamma}$ (see Subsection C) with about 8\% accuracy.

%%%%%%%%%%%%%%%%%%%%%%%%%%%%%%%%%%%%%%%%%%%%%%%%%%%%%%%%%%%%%%%%%%%%%%%%%%%%%%%%%%%%%%
%%
%%    FT
%%
%%%%%%%%%%%%
%\newpage
\subsection{The penguin form factor $F_T(0,0)$}
The form factors $F_{TA,TV}(q^2,q'^2)$ may be calculated via the $B$-meson 2DAs using HQET formula
(see e.g. \cite{offen2007}) 
\begin{eqnarray}
  \label{2BS}
  \langle 0|\bar s(x)\Gamma b(0)|\bar B_s(p)\rangle=-\frac{if_B M_B}{4}\int d\lambda e^{-i\lambda px}
\,{\rm Tr}\bigg\{ \gamma_5 \Gamma (1+\slashed{v})\bigg[\phi_+(\lambda)-\frac{\phi_+(\lambda)-\phi_-(\lambda)}{2 vx}\slashed{x}\bigg]\bigg\}
\end{eqnarray}
leading
to the following expression for the $F_{T}(0,0)=F_{TA}(0,0)=F_{TV}(0,0)$: 
\begin{eqnarray}
\label{FT}
F_T(0,0)=-Q_s f_B M_B
\int d\lambda \frac{\phi_+(\lambda)(1-\lambda)+\bar\Phi(\lambda)}{m_s^2+\lambda(1-\lambda)M_B^2}
\end{eqnarray}
where
\begin{eqnarray}
\bar \Phi(\lambda)=\int\limits_0^\lambda d\lambda' \big[\phi_{+}(\lambda')-\phi_{-}(\lambda')\big]. 
\end{eqnarray}
The absence of the kinematical singularity at $vx\to 0$ in Eq.~(\ref{2BS}) requires that the primitive
$\bar \Phi(\lambda)$ vanishes at the boundaries of the 2DA support region. Then the parts integration in $\lambda$,  
necessary to handle the $1/vx$ term, does not contain any nonzero surface term. 

All contributions $O(m_s/M_B)$ in the numerator of (\ref{FT}) are omitted; keeping such terms will be inconsistent as 
contributions of this order arise also from the diagrams describing
photon emission from the $B$-quark which are not taken into account.
According to the estimates obtained in \cite{mnk2018},
corrections due to the photon emission from the $b$-quark amount to 10-20\% of the leading
contribution (\ref{FT}). 

For the 2DAs we use the same model from \cite{braun2017} as we do for 3DAs:
\begin{eqnarray}
\label{2DAsa}
\phi_+(\lambda)&=&\frac{5}{8\omega_0^5}\lambda (2\omega_0-\lambda)^3\theta(2\omega_0-\lambda),\\
\label{2DAsb}
\phi_-(\lambda)&=&\frac{5}{192\omega_0^5}(2\omega_0-\lambda)^2\bigg[
  6(2\omega_0-\lambda)^2-\frac{7(\lambda_E^2-\lambda_H^2)}{M_B^2\omega_0^2}
 (15\lambda^2-20\omega_0\lambda+4\omega_0^2) \bigg]
\theta(2\omega_0-\lambda).
\end{eqnarray}
An explicit check shows that $\bar \Phi(\lambda=0)=\bar \Phi(2\omega_0)=0$. 
The contribution of the term $\sim(\lambda_E^2-\lambda_H^2)$ to the form factor turns
out numerically negligible and may be safely omitted. 

\begin{figure}[ht]
\begin{center} 
  \includegraphics[height=5cm]{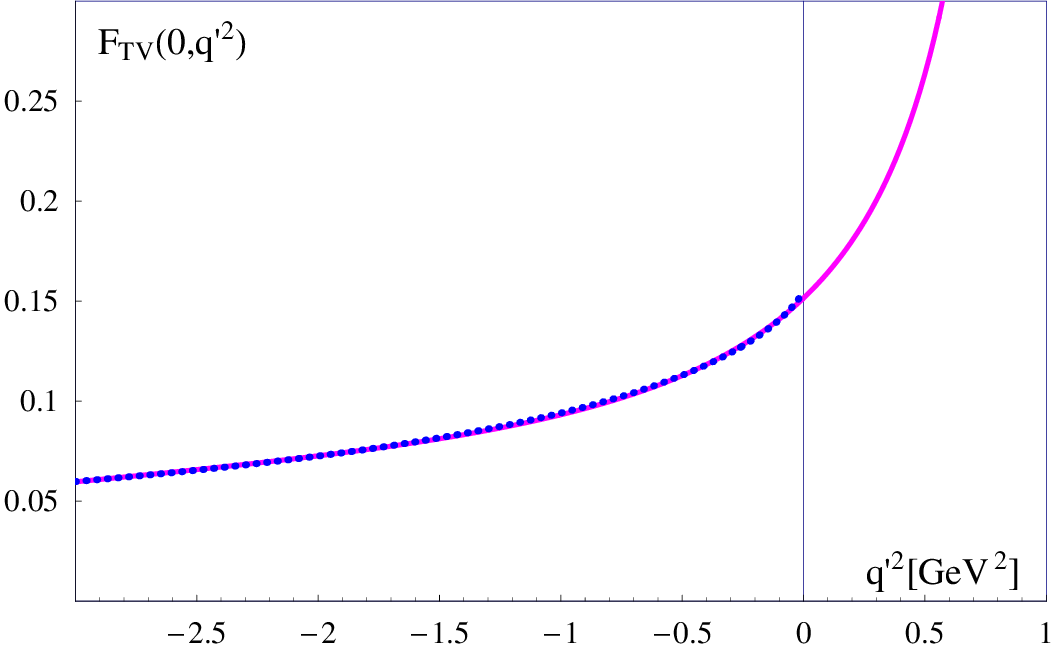} \qquad \includegraphics[height=5cm]{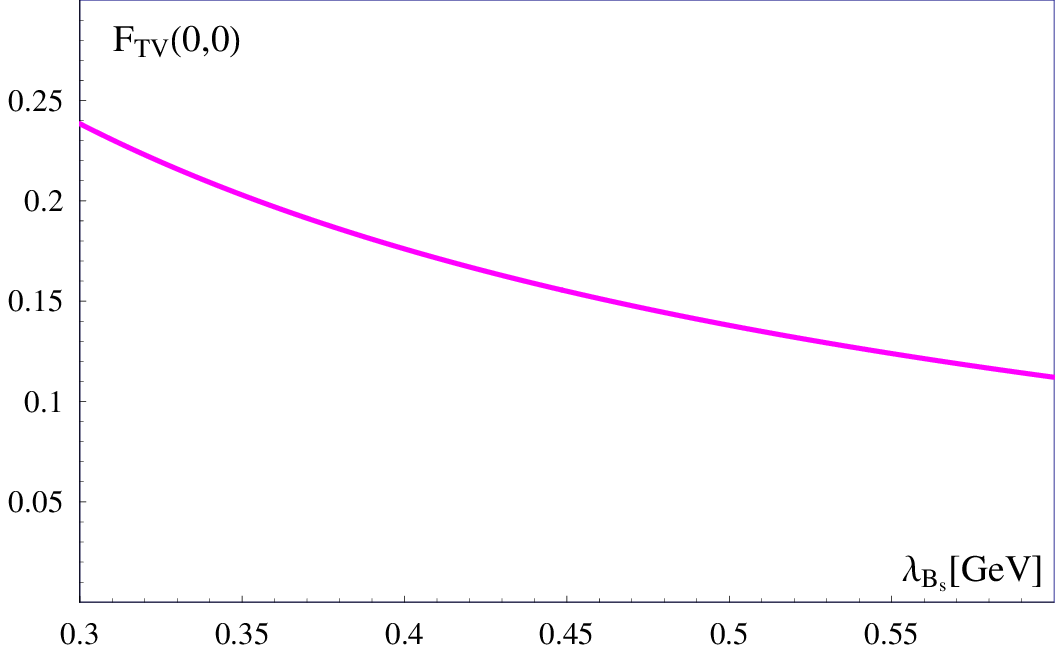} 
\caption{\label{Plot:FT}
(a) The form factor $F_{TA}(0,q'^2)$,
  where $q'$ is the momentum of the photon emitted from the $s$-quark, for 2DAs given by (\ref{2DAsa})
  and (\ref{2DAsb})
and our benchmark value $\lambda_{B_s}=0.45$ GeV.
Dotted line - calculation results; solid line - interpolation using the modified
pole formula Eq.~(\ref{fit}). 
(b) The dependence of $F_{TA}(0,0)$ on $\lambda_{B_s}$ in the range $\lambda_{B_s}=0.45\pm 0.15$ GeV.
}
\end{center}
\end{figure}
Let us obtain numerical estimates for $F_T(0,0)$.
Notice that the numerator of the integrand in Eq.~(\ref{FT}) contains factor $\lambda$,
so no singularity at $\lambda\to 0$ arises in the integrand even in the limit $m_s\to 0$.
Therefore, the $F_{TA}(0,0)$ may be calcualted by applying directly the
representation for the form factor Eq.~(\ref{FT}). [Recall that in order to obtain the form factors $H_{A,V}(0,0)$
we had to perform extrapolation from the region $q'^2\lesssim -0.5$ GeV$^2$.] 
Figure~\ref{Plot:FT}(a) shows that the $q'^2$-behaviour of $F_{TA}(0,q'^2)$ is well compatible with the
location of the physical pole at $q'^2=M_\phi^2$ in a broad range of $q'^2<0$, up to $q'^2=0$.
Figure~\ref{Plot:FT}(b) illustrates the sensitivity of $F_T(0,0)$ to $\lambda_{B_s}$. 
This dependence is not negligible and partly compensates the $\lambda_{B_s}$-dependence of $H_{A,V}(0,0)$, 
leading to more stable predictions for $\delta C_{7\gamma}$.

For our further numerical estimates we use $F_T(0,0)=0.155\pm 0.015$ for $\lambda_{B_s}=0.45$ GeV, where the uncertainty of
$\sim$10\% is assigned on the basis of the size of the neglected $1/M_B$-suppressed contributions which have been
calculated in \cite{mnk2018}.
%\newpage

\subsection{$\delta C_{7\gamma}$}
We are ready to evaluate the relative contribution of the nonfactorizable charming loops given by Eq.~(\ref{deltaC7}): 
\begin{eqnarray}
\label{deltaC7b}
\delta_{i}C_{7\gamma}=8\pi^2
\,Q_s Q_c \frac{C_2}{C_{7\gamma}m_b}\rho^{(i)}_{cc},\qquad
\rho^{(i)}_{cc}=\frac{R_{iE}\lambda_E^2+R_{iH}\lambda_H^2}{F_{T}},\qquad i=A,V.
\end{eqnarray}
This expression is useful as it allows us to separate the uncertainties related to the precise
values of the Wilson coefficients and $m_b$ to be used in the numerical estimates and those
related to the description of the $B$-meson structure.

Let us focus on $\rho^{(i)}_{cc}$. The parameters $\lambda_{B_s}$, $\lambda_E$ and $\lambda_H$, entering $\rho^{(i)}_{cc}$,
are strongly correlated with each other and should not be treated as independent quantities.
As noticed in \cite{braun2017} (see also
\cite{grozin1997,nishikawa2014}), QCD sum rules suggest an approximate relation $\lambda_H^2=2\lambda_E^2$.
Combining this relation with the constraints from the QCD equations of motion \cite{braun2017},
one obtains approximate relations
\begin{eqnarray}
\lambda_E^2\simeq 0.6 \lambda_{B_s}^2,\qquad \lambda_H^2\simeq 1.2 \lambda_{B_s}^2. 
\end{eqnarray}
As the result, $\rho^{i}_{cc}$ turns out to be the function of one variable, $\lambda_{B_s}$,
and the appropriate combinations of the form factors that determines the correction $\delta C_{7\gamma}$ are 
$R_{VE}+2R_{VH}$ and $R_{AE}+2R_{AH}$, presented in Table \ref{Table:Results}.

\begin{figure}[h!]
\begin{center} 
\includegraphics[height=5cm]{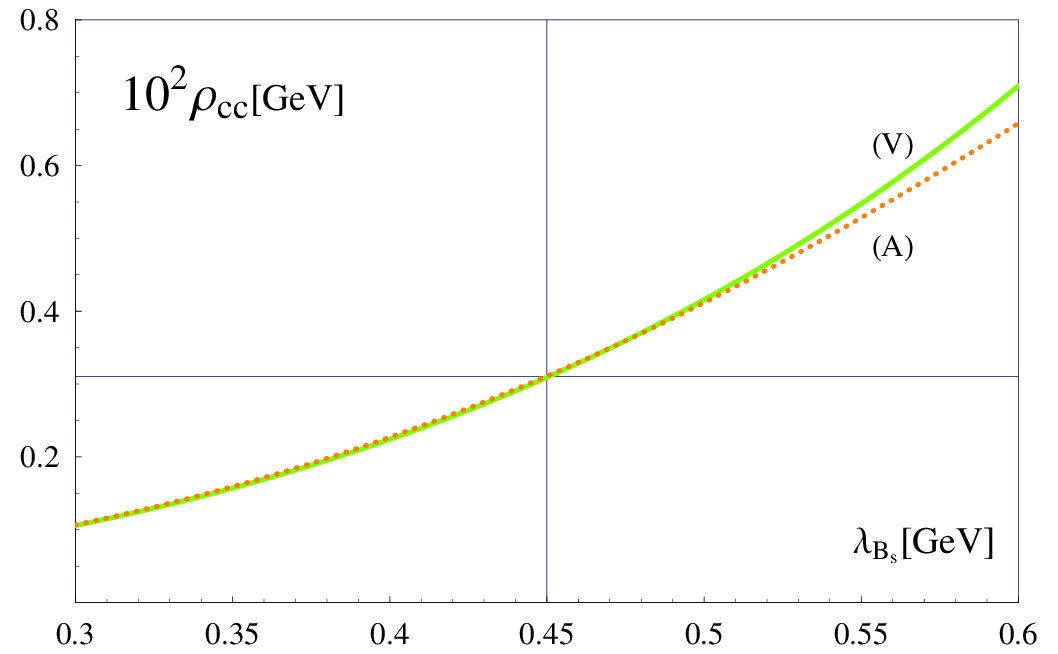} 
\caption{\label{Plot:rho}
The functions $\rho^{(i)}_{cc}, i=A,V$ [Eq.~(\ref{deltaC7b})] vs $\lambda_{B_s}$.}
\end{center}
\end{figure}
Fig.~\ref{Plot:rho} shows the dependence of $\rho^{i}_{cc}$ on $\lambda_{B_s}$ for $R_{iH,iE}$ given in
Fig.~\ref{Plot:RvslambdaBs} and $F_T$ given in Fig.~\ref{Plot:FT}.
The results presented in the plot correspond to Scenario I calculated with 3DAs of Model IIB.
To a good accuracy, we may set $\rho^{(A)}_{cc}=\rho^{(V)}_{cc}$.

Using the values $C_2=-1.1$, $C_{7\gamma}=-0.31$ and taking $\overline{m}_b(\overline{m}_b)=4.2$ GeV for $m_b$, 
Eq.~(\ref{deltaC7b}) yields: 
\begin{eqnarray}
\delta C_{7\gamma}(\lambda_{B_s})=(0.15\,{\rm GeV^{-1}})\,10^2 \rho_{cc}(\lambda_{B_s}). 
\end{eqnarray}
The final result for $\delta C_{7\gamma}$ is very sensitive to the precise value of $\lambda_{B_s}$.
Recall, however, that for a given value of $\lambda_{B_s}$, $\rho_{cc}$ may be calculated with an
accuracy around 10\%. So we prefer to present our results for $\delta C_{7\gamma}$ as the function of
$\lambda_{B_s}$. For our benchmark point $\lambda_{B_s}^0=0.45$ GeV, we find
\begin{eqnarray}
\delta C_{7\gamma}(\lambda^0_{B_s})=0.045\pm 0.004. 
\end{eqnarray}
For $\lambda_{B_s}$ in the range  $0.3<\lambda_{B_s}({\rm GeV})<0.6$, the corresponding $\delta C_{7\gamma}$
covers the range 
\begin{eqnarray}
\delta C_{7\gamma}=(2\div 10)\%. 
\end{eqnarray}
We therefore conclude that the effect of nonfactorizable charming loops is expected at the level of a few percent.
As soon as the parameter $\lambda_{B_s}$ is known with a better accuracy, our results allow one to
obtain a (relatively) accurate estimate for $\delta C_{7\gamma}$.

\newpage
 
\section{\label{sec:6} Discussion and conclusions}
We presented a detailed analysis of NF charming loops in FCNC $B_s\to\gamma\gamma$ decay and reported the following results: 

\vspace{0.2cm}
\noindent
(i) We derived and made use of the expression for the $\langle VVA\rangle$ charm-quark loop
that is fully given in terms of the gluon field strength $G_{\mu\nu}(x)$. This has an advantage that no {\it explicit} 
use of any specific gauge for the gluon field is necessary. Still, nonlocal operator describing the charm-loop contribution
to the amplitude of FCNC $B$-decay contains field operators at different coordinates, $\bar s(y) G_{\mu\nu}(x) b(0)$, and thus
needs the Wilson lines joining the field operators at different points. These Wilson lines are reduced to unity operators
in the Fock-Schwinger gauge $x_\mu A_\mu(x)=0$, so this gauge is used {\it implicitly}. 

\vspace{0.2cm}
\noindent
(ii) We studied the generic non-collinear 3BS of the $B$-meson; this quantity
contains new Lorentz structures and new 3DAs compared to collinear and double-collinear 3BS. 
We took into account constraints on the non-collinear 3BS coming from the requirement of 
analyticity and continuity \cite{m2023} and implemented proper modifications
of the corresponding 3DAs $X_i(\omega,\lambda)$ at large values of their arguments. 

\vspace{0.2cm}
\noindent
(iii) We calculated the form factors $H_i(q^2=0,q'^2=0)$, $i=A,V$, describing the $B\to\gamma\gamma$ amplitude.
Whereas $q^2=0$ [$q$ the momentum emitted from the charm-quark loop] may be set directly in the analytic formulas,
the physical point $q'^2=0$ [$q'$ the momentum emitted by the light $s$-quark] was reached by an extrapolation
from the spacelike region $-5\le q'^2({\rm GeV}^2)\le -(0.5\div 1)$.
An explicit dependence of the form factors and the correction $\delta C_{7\gamma}$ on the parameter $\lambda_{B_s}$
was calculated. Taking into account
all uncertainties (excluding that of $\lambda_{B_s}$), we found that for any specific value
of $\lambda_{B_s}$, $\delta C_{7\gamma}$ may be obtained with better than 8\% accuracy.
For our benchmark point  $\lambda^0_{B_s}=0.45$ GeV, we found
\begin{eqnarray}
\delta C_{7\gamma}(\lambda^0_{B_s})=0.045\pm 0.004.\nonumber
\end{eqnarray}
For $\lambda_{B_s}$ in the range
$0.3 < \lambda_{B_s}({\rm GeV}) < 0.6$, $\delta C_{7\gamma}$ covers the range 2-10\%.
Thus, one should expect the  NF charm-loop correction to $B\to\gamma\gamma$ decays at the level of a few \%. 
As soon as a more accurate value of $\lambda_{B_s}$ is available, our results allow one to obtain
the corresponding $\delta C_{7\gamma}$. 

\vspace{0.2cm}
\noindent
(iv) In the double-collinear kinematics that dominates the NF charm-loop contribution to FCNC $B$-decay amplitudes
in the HQ limit \cite{m2023}, the leading contribution to the amplitude is given by the
convolution of the form factor $F_0$ describing the charm-quark loop Eq.~(\ref{cquarkloop}) and the following combination
of the 3DAs \cite{wang2022,wang2022b}: 
\begin{eqnarray}
\label{wang}
\Psi_A+\Psi_V+2(W^{(y)}+Y_A^{(y)}-\tilde{Y}_A^{(y)})\sim (\lambda_E^2+\lambda_H^2). 
\end{eqnarray}
[The explicit expressions for these 3DAs show that this combination is proportional to $(\lambda_E^2+\lambda_H^2)$.] 
Eq.~(\ref{wang}) implies that in the HQ limit, $R_{VE}=R_{VH}$ and $R_{AE}=R_{AH}$.
Our results presented in Fig.~\ref{Plot:H} show that these relations are sizeably violated by $O(\lambda_{B_s}/M_B)$
corrections that come into the game via other Lorentz structures describing the charm-quark loop (\ref{cquarkloop})
and other 3DAs and worth to be taken into account. Numerically, these corrections are around 20\%. 

%\vspace{.2cm}
It might be useful to notice that the $B_s\to\gamma\gamma$ decay amplitude receives contributions from the
weak-annihilation type diagrams \cite{wa1,wa2,wa3}. The weak-annihilation mechanism differs very much from
the mechanism discussed in this paper and is therefore beyond the scope of our interest here. However,
weak-annihilation diagrams should be taken into account in a complete analysis of $B_s\to\gamma\gamma$ decays. 

\vspace{.2cm}
In conclusion, we emphasize that the approach of this paper may be readily applied to the analysis of nonfactorizable
charming loops in other FCNC $B$-decays and looks promising for treating NF effects in nonleptonic $B$-decays
(see e.g. \cite{piscopo2023}).

\vspace{.5cm}\noindent 
{\it Acknowledgments.}
We are pleased to express our gratitude to M.~Ferre, E.~Kou, O.~Nachtmann, and H.~Sazdjian for a fruitful
collaboration at the early stage of this project, and to Q.~Qin, Y.-L.~Shen, C.~Wang and Y.-M.~Wang
for interesting discussions. Our special thanks and due to Yu-Ming Wang for his illuminating remarks and comments.
D.~M. gratefully acknowledges participation at the Erwin Schr\"odinger Institute (ESI) thematic program
``Quantum Field Theory at the Frontiers of the Strong Interaction'' which promoted a deeper understanding
of the problems discussed and influenced much the final form of the paper.
The research was carried out within the framework of the program ``Particle Physics and Cosmology'' 
of the National Center for Physics and Mathematics.

\newpage
%\newpage
\appendix
\renewcommand\theequation{A.\arabic{equation}}
\section{Comparison with the charm-loop contribution from \cite{mnk2018}}
\label{sec:appA}
\renewcommand\theequation{A.\arabic{equation}}
Here we compare the definition of the $B\to\gamma\gamma$ amplitude (\ref{Acharm1}) with the definition adopted
in \cite{mnk2018}. The starting point in \cite{mnk2018} is the matrix element 
\begin{eqnarray}
H_{\rho\eta}(q,q')=i\int dz e^{i q z}\langle0|T\{\bar c\gamma_\rho c(z),j^{\rm e.m.}_\eta(0)|B_s(p)\rangle, \qquad p=q+q', 
\end{eqnarray}
where quark operators are Heisenberg operators in the SM, i.e. the corresponding $S$-matrix includes weak and strong interactions.
For real photons in the final state, at least one soft gluon should be emitted from the
charm-quark loop, we expand the $S$-matrix to first order in $G_F$ and first order in $g_s$: 
\begin{eqnarray}
\label{A3}
H_{\rho\eta}(q',q) = i \int d z e^{i q z}
\langle 0|T\lbrace\bar c(z)\gamma_{\rho}c(z), i\int dy\, L_{\text{weak}}(y), i\int dx\, L_{Gcc}(x),
eQ_s\,\bar s(0)\gamma_{\eta}s(0)\rbrace| B_s(p)\rangle.
\end{eqnarray}
For comparison with Eq.~(\ref{Acharm1}), we place $L_{\rm weak}$ at $y=0$ by shifting coordinates
of all operators through the translation
${\cal O} (x)= e^{i\hat P y} {\cal O}(x-y) e^{-iy\hat P}$.  
Using the relations 
$\langle 0|e^{i(\hat P y)} = \langle 0|$ and $e^{-i(y\hat P)} | B_s(p)\rangle=e^{-i(py)} | B_s(p)\rangle$,
and changing the variables $x-y\to x$, $z-y\to z$, $y\to -y$, we find 
\begin{eqnarray}
  \label{A4}
H_{\rho\eta}(q,q') = i^3\, e\,Q_s\int dx dy dz \: e^{i q z+iq'y}
\langle 0|T\lbrace\bar c(z)\gamma_{\rho}c(z),  L_{\text{weak}}(0),
L_{Gcc}(x), \bar s(y)\gamma_{\eta}s(y)\rbrace| B_s(p)\rangle.  
\end{eqnarray}
Taking into account that $L_{\rm weak}=-H^{b\to s\bar cc[8\times8]}_{\rm weak}$ [the latter given by Eq.~(\ref{Heff8times8})],
we obtain 
\begin{eqnarray}
\label{A5}
H_{\rho\eta}(q,q')& =& 
2C_2\,\frac{G_F}{\sqrt{2}} V_{cb}V^*_{cs}e\,Q_s\, i\int dx dy d z\: e^{i q z+iq'y}
\langle 0 |T\lbrace\bar c(z)\gamma_{\rho}c(z), \bar c(0) \gamma_{\mu}(1-\gamma_5)t^a c(0), \bar c(x) \gamma_\nu t^b c(x)|0 \rangle
\nonumber\\
&&\hspace{4cm}\times
\langle 0|T\left\{\bar s(y)\gamma_{\eta}s(y),\bar s(0)\gamma_{\mu}(1-\gamma_5)t^a b(0),B^b_\nu(x)\right\}| B_s(p)\rangle
\end{eqnarray}
It is convenient to insert under the integral (\ref{A5}) the identity
\begin{eqnarray}
  \label{A6}
B^b_\nu(x)=  \frac{1}{(2\pi)^4} \int d\kappa dx' B^b_\nu(x' ) e^{i\kappa (x-x')}.
\end{eqnarray}
This allows us to isolate the contribution of the charm-quark loop $\Gamma^{\mu\nu\rho(ab)}_{cc}(\kappa ,q)$: 
\begin{eqnarray}
\label{A7}
H_{\rho\eta}(q,q')& =& 2C_2\,\frac{G_F}{\sqrt{2}} V_{cb}V^*_{cs} e Q_s\frac{i}{(2\pi)^4}\int  dy \, e^{iq'y} \;
d \kappa e^{-i\kappa x'}
\langle 0|T\lbrace\bar s(0)\gamma_{\mu}(1-\gamma_5) b(0), B^b_\nu(x'), \bar s(y)\gamma_{\eta}s(y)\rbrace| B_s(p)\rangle
\nonumber
\\
&&\times\underbrace{
  \int dx dz \, e^{i q z+i \kappa x}
  \langle 0 |T\lbrace\bar c(z)\gamma_{\rho}c(z),\bar c(0) \gamma_{\mu} (1-\gamma_5)t^a c(0),
  \bar c(x) \gamma_\nu t^b  c(x)\rbrace|0 \rangle
}_{\Gamma^{\mu\nu\rho(ab)}_{cc}(\kappa ,q)}. 
\end{eqnarray} 
Using momentum representation for the $s$-quark propagator
\begin{eqnarray}
%  \overbracket{ s(y)\bar s}(0)
\langle 0|T\{s(y)\bar s(0) \}|0\rangle =\frac{1}{(2\pi)^4i}\int dk\, e^{-iky}\frac{\slashed{k}+m_s}{m_s^2-k^2-i0},
\end{eqnarray}
we obtain 
\begin{eqnarray}
  H_{\rho\eta}(q,q')&=&2C_2\,\frac{G_F}{\sqrt{2}} V_{cb}V^*_{cs}e Q_s\\
&&\times\underbrace{\frac{1}{(2\pi)^8}\int{dk}dy e^{-i(k-q')y} dx d\kappa  e^{-i\kappa x}
\Gamma_{cc}^{\mu\nu\rho(ab)}(\kappa ,q)
\langle 0|\bar s(y)\gamma^\eta\frac{\slashed  k+m_s}{m_s^2-k^2}\gamma^\mu(1-\gamma^5)t^a B^b_{\nu}(x)b(0)|B_s(p)\rangle}_{A_{\rho\eta}}.\nonumber
\end{eqnarray}
Comparing this expression with Eq.~(\ref{Acharm3a}) gives a useful relation 
\begin{eqnarray}
A^{B\to\gamma\gamma}_{\rm charm}=2eQ_c\,H_{\rho\eta}\,\varepsilon_\rho\varepsilon'_\eta. 
\end{eqnarray} 
The amplitude $H_{\rho\eta}$ may be decomposed via the form factors
$\tilde H_{A,V}$ (denoted as $H_{A,V}$ in Eq.~(5.2) from \cite{mnk2018}): 
\begin{eqnarray}
H_{\rho\eta}(q,q')=-\frac{G_F}{\sqrt2}V_{cb}V^*_{cs}e
\Big[
\epsilon_{\rho\eta qq'}\tilde H_V-i \left(g_{\rho\eta} q'q-q'_\rho q_\eta\right) \tilde H_V \Big]. 
\end{eqnarray}
Comparing with (\ref{Acharm3b}) we find
\begin{eqnarray}
\tilde H_{V(A)}=-2C_2 Q_s H_{V(A)}.
\end{eqnarray}
\newpage
\renewcommand\theequation{B.\arabic{equation}}
\section{Anathomy of the form factors $H_{A,V}(0,q_0'^2)$ at $q_0'^2=-2$ GeV$^2$.\label{Appendix_Anathomy}}
This Appendix presents detailed numerical results for the form factors at $q^2=0$ and $q'^2=q'^2_0=-2$ GeV$^2$,
giving separately contributions of different Lorentz 3DAs for the coefficients $R_{iE}$ and $R_{iH}$
defined in Eq.~(\ref{Ri}). 
Table~\ref{Table:ffsI} gives the results obtained with Scenario I of Section \ref{Section4C} for 3DAs
given by Eqs.~(\ref{phi3})--(\ref{phi6}) [Model IIB of Eq. (5.27) from \cite{braun2017}].
Table \ref{Table:ffsII} presents the results obtained with  Scenario II and the same set of 3DAs. 
Table~\ref{Table:ffsIprime} gives the results corresponding to Scenario I but using different
3DAs, those of Model IIA of Eq. (5.23) from \cite{braun2017}. 
Separate contributions coming from different Lorentz 3DAs are isolated keeping the explicit
dependence on $w_x$ and $w_y$, $w_x+w_y=1$. The results correspond to  $\lambda_{B_s}=0.45$ GeV
and the central values of $m_c$ and $m_s$. For 3DAs modified according to Scenario I, $a=0.05$.

\begin{table}[ht]
\centering 
\caption{\label{Table:ffsI}
$H_V(0,q_0'^2)$ and $H_A(0,q_0'^2)$ at $q_0'^2=-2$ GeV$^2$
calculated for Scenario I and 3DAs of Eqs.~(\ref{phi3})--(\ref{phi6}) for $\lambda_{B_s}=0.45$ GeV and $a=0.05$. 
$H_{i}(0,q'^2)=R_{iE}\lambda_E^2+R_{iH}\lambda_H^2, i=A,V$.}
\begin{tabular}{|c||c|c||c|c|}
  \hline
  &  \multicolumn{2}{|c||}{ $10^5 H_V(q^2,q_0'^2)$ } & \multicolumn{2}{c|}{ $10^5 H_A(q^2,q_0'^2)$ }\\
  \hline
      {Lorentz 3DA} & $10^5R_{VE}$ [GeV$^{-1}$] & $10^5R_{VH}$ [GeV$^{-1}$] &$10^5R_{AE}$ [GeV$^{-1}$] & $10^5R_{AH}$ [GeV$^{-1}$] \\
 \hline
 $\psi_A$          &  45.84 & 7.54 & 43.44 & 7.67 \\
 $\psi_V$          &  4.62 & 25.2 & 4.61 & 26.7 \\
 ${X}_A$
& $-0.28\,w_x + 0.49\,w_y$
 & $-1.64\,w_x + 0.99\,w_y$
 & $-0.39\,w_x + 0.24\,w_y$
 & $-1.63\,w_x + 0.81\,w_y$
 \\
 ${Y}_A$
 & $-8.37\,w_x +1.52\,w_y$
 & $8.33\,w_x +0.06\,w_y$
 & $-8.52\,w_x +0.70\,w_y$
 &  $8.48\,w_x + 0.63\,w_y$
 \\ 
 ${\tilde{X}}_A$
 & $-0.65\,w_x - 0.05\,w_y$
 & $-0.11\,w_x + 0.11\,w_y$
 & $-0.70\,w_x + 0.22\,w_y$
 & $-0.07\,w_x - 0.19\,w_y$
 \\
 ${\tilde{Y}}_A$
 & $1.65\,w_x - 0.01\,w_y$
 & $-2.45\,w_x - 1.51\,w_y$
 & $1.70\,w_x - 0.59\,w_y$
 & $-2.51\,w_x - 0.68\,w_y$
 \\
 ${W}$ 
 &  $-0.61\,w_x - 0.39\,w_y$
 &  $-0.61\,w_x - 0.39\,w_y$
 &  $-0.63\,w_x - 0.35\,w_y$
 &  $-0.63\,w_x - 0.35\,w_y$
 \\
 ${Z}$
 &  $0.93\,w_x + 0.27\,w_y$
 &  $-0.31\,w_x -0.19\,w_y$
 &  $0.95\,w_x + 0.03\,w_y$
 &  $-0.31\,w_x -0.02\,w_y$
 \\
\hline
\end{tabular}
%}
\end{table}
Setting $w_x=\frac12(1+\delta w)$, $w_y=\frac12(1-\delta w)$ and summing the contribuons of all 3DAs in Table~\ref{Table:ffsI},
we obtain,
\begin{eqnarray}
  H_V(0,q_0'^2)=(47.72 - 4.58 \,\delta_w)\mbox{ [GeV$^{-1}$]} \lambda_E^2
  + (33.88 + 2.07 \, \delta_w)\mbox{ [GeV$^{-1}$]} \lambda_H^2. 
\end{eqnarray}
For the expected relationship $\lambda_H^2=2\lambda_E^2$, the dependence on $\delta w$ is further suppressed
leading to extremely stable predicitons with respect to $\delta_w$:
\begin{eqnarray}
H_V(0,q_0'^2)=(115.48 - 0.44\, \delta_w)\mbox{ [GeV$^{-1}$]} \lambda_E^2. 
\end{eqnarray}

%%%%%%%%%%%%%%%%%%%%%%%%%%%%%%%%%%%%%%%%%%%%%%%%%%%%%%%%%%%%%%%%%%%%%%%%%%%%%%%%%
\begin{table}[ht]
\centering
\caption{$H_V(0,q_0'^2)$ and $H_A(0,q_0'^2)$ at $q_0'^2=-2$ GeV$^2$
  calculated for Scenario II and 3DAs of Eqs.~(\ref{phi3})--(\ref{phi6}); $\lambda_{B_s}=0.45$ GeV. }
   \label{Table:ffsII}
   \vspace{1ex}
\begin{tabular}{|c||c|c||c|c|}
  \hline
  &  \multicolumn{2}{|c||}{ $10^5 H_V(q^2,q_0'^2)$ } & \multicolumn{2}{c|}{ $10^5 H_A(q^2,q_0'^2)$ }\\
  \hline
      {Lorentz 3DA} & $10^5R_{VE}$ [GeV$^{-1}$] & $10^5R_{VH}$ [GeV$^{-1}$] & $10^5R_{AE}$ [GeV$^{-1}$] & $10^5R_{AH}$ [GeV$^{-1}$] \\
 \hline
 $\psi_A$          &  45.84 & 7.54 & 43.44 & 7.67 \\
 $\psi_V$          &  4.62 & 25.2 & 4.61 & 26.7 \\
 ${X}_A$
 & $-1.54\,w_x - 1.17\,w_y$
 & $-4.08\,w_x - 1.71\,w_y$
 & $-1.02\,w_x - 1.40\,w_y$
 & $-4.73\,w_x + 0.11\,w_y$
 \\
 ${Y}_A$
 & $-17.4\,w_x -5.93\,w_y$
 & $25.3\,w_x - 11.4\,w_y$
 & $-17.6\,w_x + 5.36\,w_y$
 &  $25.4\,w_x - 12.0\,w_y$
 \\
 ${\tilde{X}}_A$
 & $-1.73\,w_x - 0.80\,w_y$
 & $-0.12\,w_x + 0.32\,w_y$
 & $-1.62\,w_x + 0.59\,w_y$
 & $-0.42\,w_x - 0.65\,w_y$
 \\
 ${\tilde{Y}}_A$
 & $3.53\,w_x + 10.9\,w_y$
 & $-4.89\,w_x - 5.69\,w_y$
 & $3.51\,w_x + 11.5\,w_y$
 & $-4.96\,w_x - 5.11\,w_y$
 \\
 ${W}$
 &  $2.27\,w_x - 0.52\,w_y$
 &  $2.27\,w_x - 0.52\,w_y$
 &  $2.38\,w_x - 0.38\,w_y$
 &  $2.38\,w_x - 0.38\,w_y$          
 \\
 ${Z}$
 &  $6.45\,w_x + 0.75\,w_y$
 &  $-0.96\,w_x -0.50\,w_y$
 &  $6.28\,w_x + 0.05\,w_y$
 &  $-0.93\,w_x -0.02\,w_y$
 \\
\hline
\end{tabular}
\end{table}
For Scenario II ( Table~\ref{Table:ffsII}), one finds 
\begin{eqnarray}
H_V(0,q_0'^2)=(53.81 - 11.76\,\delta_w)\mbox{ [GeV$^{-1}$]} \lambda_E^2 + (31.75 + 18.51 \, \delta_w)\mbox{ [GeV$^{-1}$]} \lambda_H^2. 
\end{eqnarray}
For $\lambda_H^2=2\lambda_E^2$, this gives 
\begin{eqnarray}
H_V(0,q_0'^2)=(117.31 + 25.25\, \delta_w)\mbox{ [GeV$^{-1}$]} \lambda_E^2. 
\end{eqnarray}
For $\delta_w=0$, both Scenario I and II give close results but the sensitivity of the form factor 
to $\delta_w$ in Scenario II is strong.

%%%%%%%%%%%%%%%%%%%%%%%%%%%%%%%%%%%%%%%%%%%%%%%%%%%%%%%%%%%%%%%%%%%%%%%%%%%%%%%%%%%%%%%%%%%%%%%%%%%
%For further comparison, we provide the results obtained with Scenario I but for slightly different set of 

\begin{table}[ht]
\centering 
\caption{\label{Table:ffsIprime}
$H_V(0,q_0'^2)$ and $H_A(0,q_0'^2)$ at $q_0'^2=-2$ GeV$^2$
  calculated for Scenario I and 3DAs of Model IIA given by Eq.~(5.23)
  of \cite{braun2017} for $\lambda_{B_s}=0.45$ GeV and $a=0.05$. }
\begin{tabular}{|c||c|c||c|c|}
  \hline
  &  \multicolumn{2}{|c||}{ $10^5 H_V(q^2,q_0'^2)$ } & \multicolumn{2}{c|}{ $10^5 H_A(q^2,q_0'^2)$ }\\
  \hline
      {Lorentz 3DA} & $10^5R_{VE}$ [GeV$^{-1}$] & $10^5R_{VH}$ [GeV$^{-1}$] &$10^5R_{AE}$ [GeV$^{-1}$] & $10^5R_{AH}$ [GeV$^{-1}$] \\
 \hline
 $\psi_A$          &  46.7 & 6.97 & 44.8 & 7.05 \\
 $\psi_V$          &  4.74 & 28.9 & 4.74 & 30.1 \\
 ${X}_A$
 & $0.08\,w_x + 0.00\,w_y$
 & $-0.39\,w_x + 0.17\,w_y$
 & $0.07\,w_x - 0.04\,w_y$
 & $-0.39\,w_x + 0.18\,w_y$
 \\
 ${Y}_A$
 & $-3.50\,w_x + 0.96\,w_y$
 & $1.79\,w_x - 0.31\,w_y$
 & $-3.56\,w_x + 0.70\,w_y$
 & $1.82\,w_x - 0.21\,w_y$
 \\
 ${\tilde{X}}_A$
 & $-0.17\,w_x - 0.02\,w_y$
 & $0.04\,w_x + 0.03\,w_y$
 & $-0.18\,w_x + 0.06\,w_y$
 & $0.05\,w_x - 0.05\,w_y$
 \\
 $\tilde{Y}_A$
 & $0.52\,w_x + 0.31\,w_y$
 & $-1.18\,w_x - 0.94\,w_y$
 & $0.54\,w_x + 0.20\,w_y$
 & $-1.21\,w_x - 0.68\,w_y$
 \\
 ${W}$
 &  $-0.04\,w_x - 0.04\,w_y$
 &  $-0.04\,w_x - 0.04\,w_y$
 &  $-0.04\,w_x - 0.04\,w_y$
 &  $-0.04\,w_x - 0.04\,w_y$          
 \\
 ${Z}$
 &  $0.13\,w_x +0.05\,w_y$
 &  $-0.08\,w_x - 0.05\,w_y$
 &  $0.14\,w_x + 0.01\,w_y$
 &  $-0.08\,w_x - 0.01\,w_y$
 \\
\hline
\end{tabular}
%}
\end{table}
For Scenario I but using different 3DAs, those of Model IIA of Eq. (5.23) from \cite{braun2017}, one finds
(Table~\ref{Table:ffsIprime}): 
\begin{eqnarray}
H_V(0,q_0'^2)=(50.58 - 2.12 \,\delta_w)\mbox{ [GeV$^{-1}$]} \lambda_E^2 + (35.37 + 0.64 \, \delta_w)\mbox{ [GeV$^{-1}$]} \lambda_H^2. 
\end{eqnarray}
For $\lambda_H^2=2\lambda_E^2$, this yields:
\begin{eqnarray}
H_V(0,q_0'^2)=(121.32 - 0.84\, \delta_w)\mbox{ [GeV$^{-1}$]} \lambda_E^2. 
\end{eqnarray}

%\newpage

\end{document}